\documentclass[aps,pre,reprint,twocolumn,showpacs,superscriptaddress]{revtex4-2}
\usepackage{epsfig}
\usepackage{natbib}
\usepackage{ulem}
\usepackage{xspace}
\usepackage{xcolor}
\usepackage{graphicx}
\usepackage{amssymb}
\usepackage{cancel}
\usepackage{ulem}
\usepackage{lineno}
\usepackage{amsmath}
\usepackage{stackrel}
\usepackage{appendix}
\usepackage{url}
 \usepackage{enumitem}
 \usepackage{hyperref}
\usepackage{feynmp-auto}
\usepackage{tikz}
\usepackage{bm}
\usepackage{mathrsfs}

 \usepackage{hyperref}

\hypersetup{
	colorlinks=true,
	linkcolor=blue,
	citecolor=blue,
	urlcolor=blue
}

\usepackage{color}

\definecolor{Blue}{rgb}{0.00, 0.00, 0.80}
\definecolor{Red}{rgb}{0.80, 0.00, 0.00}
\definecolor{Green}{rgb}{0.00, 0.50, 0.00}

\newcommand{\golden}{\varphi}

\newcommand{\new}[1]{{#1}}

\begin{document}

\title{Remarkable similarities in distributions of dynamical observables in chaotic systems}

\author{Lucianno Defaveri}
\affiliation{Racah Institute of Physics, Hebrew University of Jerusalem, Jerusalem 91904, Israel}
\author{Naftali R. Smith}
\affiliation{Racah Institute of Physics, Hebrew University of Jerusalem, Jerusalem 91904, Israel}

\begin{abstract}
The study of chaotic systems, where rare events play a pivotal role, is essential for understanding complex dynamics due to their sensitivity to initial conditions. Recently, tools from large deviation theory, typically applied in the context of stochastic processes, have been used in the study of chaotic systems.  Here, we study dynamical observables, $A = \sum_{n=1}^N g(\bm{x}_n)$, defined along a chaotic trajectory $\{\bm{x}_1, \bm{x}_2, \ldots, \bm{x}_N\}$. For most choices of $g(\bm{x})$, $A$ satisfies a central limit theorem: At large sequence size $N \gg 1$, typical fluctuations of $A$ follow a Gaussian distribution with a variance that scales linearly with $N$. Large deviations of $A$ are usually described by the large deviation principle, that is, $P(A) \sim e^{- N I(A/N)}$, where $I(a)$ is the rate function. 
We find that certain dynamical observables exhibit a remarkable statistical similarity: even when constructed with distinct functions $g_1(\bm{x})$ and $g_2(\bm{x})$, different observables are described by the same rate function.
We provide a physical interpretation for this striking similarity by showing that $g_1(\bm{x})-g_2(\bm{x})$ belongs to a class of functions that we call ``derived''.
Furthermore, we show that if $g(\bm{x})$  itself is ``derived'', then the distribution of $A$ becomes independent of $N$ in the large-$N$ limit, and is generally non-Gaussian (although it is mirror-symmetric). We demonstrate that \new{the position observable for certain open maps, used to model random walks and} the finite-time Lyapunov exponent (FTLE) for the logistic map \new{are} of this derived form, thus providing a simple explanation for some existing results.
\end{abstract}

\maketitle

\section{Introduction}

Dynamics of deterministic, natural, classical systems are often observed to be chaotic. This means that it is practically impossible to predict their behavior at long times, due to a strong sensitivity to the initial conditions. In particular, the distance in phase space between trajectories with similar initial conditions grows exponentially in time (the ``butterfly effect''). Chaotic dynamics can be observed in several contexts, ranging from climate systems \cite{DeCruz2018}, finance \cite{Klioutchnikov2017}, ecology \cite{Roy2019, Roy2020, Pearce2020}, and pandemic models \cite{Postavaru2021}. Often, these types of dynamics can be modeled effectively as stochastic processes, providing a description that lies at the foundations of statistical mechanics and allows us to make probabilistic predictions \cite{Geisel1982, Schell1982, Fujisaka1982, Eckmann1986, Sato2019, Albers2022}.

Large deviations (or rare events) in chaotic systems are of practical importance as they may have significant consequences. Examples include extreme weather events such as heatwaves, cold spells, floods, or droughts \cite{Wouters2016, Ragone2018, Ragone2021, Galfi2021, Galfi2021a, Shen, DErrico2022, Ashkenazy2024}. It is, however, very difficult to obtain a full theoretical understanding of even the simplest weather models due to the technical challenges involved, and one usually resorts to semi-analytic or numerical methods.
In order to make analytic progress, it is useful to consider simpler models for chaotic systems. Indeed, large deviations of dynamical (or time-integrated) observables in discrete-time chaotic systems have been studied for some time \cite{Grassberger1985,Prasad1999,Anteneodo2004,Nakao2006,Lai2007} and have recently attracted renewed interest \cite{Smith2022, Gutierrez2023, Monthus2024, Lippolis2024, Monthus2025,Gonzalez-Koda2025}.
In particular, these studies enable one to go beyond the statistical mechanics treatment of approximating the chaotic system as a white noise (at long timescales). This approximation usually correctly captures the typical fluctuations regime, in which one observes a Gaussian distribution whose variance is (approximately) proportional to the duration of the observation window.

The Donsker-Varadhan theory, which was recently adapted to the case of deterministic, discrete-time chaotic dynamics \cite{Smith2022}, yields the rate function that describes the large-deviations regime. The latter function is found by solving an auxiliary problem of finding the largest eigenvalue of a modified, or tilted, generator of the dynamics. This procedure was carried out successfully in a few cases \cite{Smith2022, Gutierrez2023, Monthus2024},
but a more complete understanding, and in particular, of the dependence of the rate function on the observable in question, is lacking.

A particular dynamical observable, the finite-time Lyapunov exponent (FTLE), is of interest as it quantifies the rate at which trajectories with similar initial conditions diverge in time. In the limit of an infinite observation window, the FTLE converges to the Lyapunov exponent of the dynamics. However, if the observation window is long but finite, fluctuations are possible and they are interesting to quantify. Indeed, the distribution of the FTLE has been studied in several previous works \cite{Grassberger1985,Prasad1999,Haller2001,Anteneodo2004,Nakao2006,Bec2006,Lai2007,Biferale2014, Gutierrez2023}.
In the particular case where the chaotic dynamics is generated by the logistic map, the FTLE distribution was found to possess several interesting properties.
Its variance decays anomalously with the duration of the observation window. Typical fluctuations follow a distribution that is not Gaussian, but is approximately mirror-symmetric (around its mean).
Furthermore, the rate function that describes large deviations has a singularity that may be interpreted as a first-order dynamical phase transition \cite{Anteneodo2004, Gutierrez2023}. Although these properties may be obtained via an exact solution \cite{Anteneodo2004}, to our knowledge, there is no intuitive understanding of an underlying physical mechanism that gives rise to these properties in this particular case, or a method of predicting whether some of them hold in other cases.

In this paper, our main aim is to study the dependence of the distribution of dynamical observables (and in particular, of the rate functions that describe their large deviations) on the observable in question. We uncover a remarkable similarity: It turns out that different observables may be described by the exact same rate function. 
Furthermore, we find a class of observables for which the distribution of both typical fluctuations and large deviations is anomalous. We call these observables ``derived''. In particular, we find that the distribution of derived observables (and in particular, its variance) becomes independent of the duration of the observation window, and that its typical fluctuations are not Gaussian but do possess an (approximate) mirror symmetry.
We show that the FTLE for the logistic map, properly shifted and scaled, is a derived observable, thus providing a more complete understanding of previous results for this particular case \cite{Anteneodo2004}. \new{Recent exact results in deterministic Floquet circuits have uncovered similarities in large deviation functions associated with space-time patterns, such as persistent empty regions. These similarities emerge from a mechanism that is analogous to the one we derive in this manuscript~\cite{Klobas2024,DeFazio2024,Garrahan_privcom_2025}.}

The remainder of the paper is organized as follows. In Section~\ref{sec:LDT-chaos} we define the type of models and observables that we study, and recall the Green-Kubo formula and the Donsker-Varadhan theory that generally describe the typical fluctuations and large-deviations regimes, respectively.
In Section~\ref{sec:identical-rate-functions} we demonstrate two particular cases of pairs of observables whose rate functions exactly coincide, in the tent and logistic maps. We then explain the mechanism behind this coincidence and treat general dynamical observables in Section~\ref{sec:delta_g}.
In Section~\ref{sec:g-derived-function} we introduce derived observables and study their distributions, and in particular, we show that they possess the properties listed above.
\new{In Section~\ref{sec:diffusion-chaos} we model transport with an open piecewise‑linear chaotic map whose dynamics reproduce a nearest‑neighbor random walk. For a specific choice of parameters, the counts of right/left hops and the activity all share the same large‑deviation rate function. We also show that the position of these walkers is a derived observable.}
In Section~\ref{sec:FTLE} we study the FTLE in the logistic map. We show that it is a derived observable and this enables us to reproduce some of the known results for its distribution in a simple way.
Finally, we briefly discuss our main findings in Section~\ref{sec:final-remarks}.

\section{Large deviations of dynamical observables in chaotic maps \label{sec:LDT-chaos}}

In this paper, we analyze a sequence of values $\bm{x}_1, \bm{x}_2, \ldots, \bm{x}_N$, where $\bm{x}_n \in \mathbb{R}^d$, with $d$ denoting the spatial dimension. The sequence begins with an initial value, $\bm{x}_1$, which is a random variable sampled from a known probability density function (PDF) $p_1(\bm{x}_1)$. Subsequent values of this sequence are then generated deterministically via the iterative rule
\begin{equation}
    \bm{x}_{n+1} = f(\bm{x}_n) \, , \label{eq:interative-map}
\end{equation} 
where $f: \mathbb{R}^d \to \mathbb{R}^d$ represents a deterministic mapping, which is assumed to be chaotic. Thus, for a given initial value $\bm{x}_1$, the entire sequence is fully known. Alternative to Eq.\,(\ref{eq:interative-map}), we can treat the system similarly to a Markov chain, where the 
distribution of each element in the sequence $P_{n}(\bm{x}_n)$ determines that of the following element $P_{n+1}(\bm{x}_{n+1})$ through
\begin{equation}
    P_{n+1}(\bm{x}) = \int \delta( \bm{x} - f(\bm{y})) P_n(\bm{y}) d \bm{y} \, .
\end{equation}
The previous equation may be rewritten as $P_{n+1}(\bm{x}) = \mathcal{L}_\mathrm{FP} P_{n}(\bm{x})$, where the Frobenius-Perron operator (FPO) $\mathcal{L}_\mathrm{FP}$ is defined as
\begin{equation}
    \mathcal{L}_\mathrm{FP} \rho(\bm{x}) = \sum_{\bm{z}_i \in f^{-1}(\bm{x})} \frac{\rho(\bm{z}_i)}{|J_f(\bm{z}_i)|} \, ,
\end{equation}
where $J_f(\bm{x})$ is the Jacobian of the map $f(\bm{x})$ and $\bm{z}_i$ are all the points in the pre-image of $\bm{x}$ (under $f$).

A very important PDF is the invariant measure (IM) of the process, denoted by $p_S(\bm{x})$. The IM represents a stationary PDF such that if $\bm{x}$ is sampled from $p_S(\bm{x})$, the transformed value $f(\bm{x})$ is also distributed according to $p_S(\bm{x})$. In the context of the FPO, the invariant measure is an eigenfunction of $\mathcal{L}_\mathrm{FP}$, associated with the largest eigenvalue (or to be precise, the eigenvalue with maximal real part), equal to one. This property is central to understanding long-term dynamics, as $p_S(\bm{x})$ captures the large-$N$ statistical behavior of the sequence, irrespective of the initial condition $p_1(\bm{x}_1)$. More specifically, regardless of $p_1(\bm{x})$, we have that $ \lim_{N \to \infty} \mathcal{L}_\mathrm{FP}^N p_1(\bm{x}) = p_S(\bm{x})$ (assuming ergodicity). For simplicity, we will assume that the initial condition is sampled from the IM, i.e., that $p_1(\bm{x}) = p_S(\bm{x})$, \new{physically, this corresponds to assuming the system has undergone a large transient and reached a steady-state.} It follows that the distribution of all elements in the sequence is also given exactly by the IM, $p_i(\bm{x}) = p_S(\bm{x})$ for $i=1,\dots,N$.

For a given sequence, we can define a dynamical observable $A$ as \cite{Smith2022, Gutierrez2023, Monthus2024}
\begin{equation}
\label{Adef}
    A = \sum_{n=1}^N g(\bm{x}_n) \, .
\end{equation}
This definition is analogous to time-averaged observables in the context of stochastic processes \cite{Touchette2009, Nickelsen2018}. Our primary goal in this manuscript is to investigate the probability distribution associated with such a dynamical observable, $P_N(A)$, in the limit of large $N$. An important example, which we will discuss in greater detail in later sections, is the finite time Lyapunov exponent (FTLE), represented by the observable function \new{$g(\bm{x}) = \ln \sigma(\bm{x})$, where $\sigma(\bm{x})$ is the largest singular value of the Jacobian of the map $f(
\bm{x})$.}

It is usually the case that the mean and variance of $A$ are proportional to $N$ at $N \gg 1$(and more generally, all of the higher cumulants as well). In fact, for the mean, this statement is exact at any $N$ since we assumed $p_1(\bm{x}) = p_S(\bm{x})$, so
\begin{equation}
    \langle A \rangle = \sum_{n=1}^N \int p_S(\bm{x}_n) g(\bm{x}_n) d\bm{x}_n = N \langle g(\bm{x}) \rangle_{S} \, . \label{eq:mean-green-kubo}
\end{equation}
Here the angular brackets $\langle \cdots \rangle$ represent an average over the initial condition $\bm{x}_1$, while the angular brackets with subscript $S$, $\langle \cdots \rangle_{S}$, represent averaging over the IM. The variance (i.e., the second cumulant) of $A$ is written explicitly as the sum of the correlation functions over all pairs of elements in the sequence,
\begin{align}
    \langle A^2 \rangle - \langle A \rangle^2= \sum_{n=1}^{N} \sum_{\delta = 1-n}^{N-n} &\Big[  \langle g(\bm{x}_n)g(\bm{x}_{n+\delta}) \rangle - \nonumber \\ &- \langle g(\bm{x}_n) \rangle \langle g(\bm{x}_{n+\delta}) \rangle \Big]  \, . \label{eq:second-cumulang-before-average}
\end{align}
Since the individual terms of the chaotic sequence are distributed according to the IM, $\langle g(\bm{x}_n) \rangle \langle g(\bm{x}_{n+\delta}) \rangle = \langle g(x) \rangle_S^2$, for any $n$. Secondly, $\langle g(\bm{x}_n)g(\bm{x}_{n+\delta}) \rangle$ is only a function of $\delta$, where $\delta = 0$ is the maximal value $\langle g(\bm{x})^2 \rangle_S$ and $\delta \to \infty$ leads to the uncorrelated limit value $\langle g(x) \rangle_S^2$.
This leads to the conclusion that the summation over $\delta$ in Eq.~(\ref{eq:second-cumulang-before-average}) is of $O(1)$. In the limit of $N\to \infty$, Eq.~(\ref{eq:second-cumulang-before-average}) therefore becomes 
\begin{equation}
    \langle A^2 \rangle - \langle A \rangle^2 \simeq N \sum_{\delta = -\infty}^{\infty} \Big[  \langle g(\bm{x}_0)g(\bm{x}_{\delta}) \rangle_S - \langle g(\bm{x}) \rangle_S^2 \Big]  \, , \label{eq:variance-green-kubo}
\end{equation}
which is the Green-Kubo formula. Note that for the purpose of Eq.~\eqref{eq:variance-green-kubo}, we think of the sequence $\bm{x}_1, \dots, \bm{x}_N$ as extended infinitely in both directions, i.e., $\dots, \bm{x}_{-1}, \bm{x}_0, \bm{x}_1, \dots$.

In the usual case, typical fluctuations of $A$ satisfy a central limit theorem with mean and variance given by Eqs.~\eqref{eq:mean-green-kubo} and \eqref{eq:variance-green-kubo}, respectively.
In order to give a more complete description of the statistics of fluctuations of $A$, that is, one that includes the contribution from rare events, we will make use of the rate function $I(a)$. This function describes the large deviations of $A$ through the large-deviations principle
\begin{equation}
    P_N(A) \sim e^{-NI(A/N)} \, , \label{eq:LDP}
\end{equation}
and it is defined formally as \cite{Touchette2009,Smith2022,Gutierrez2023,Monthus2024}
\begin{equation}
    I(a \equiv A/N) = - \lim_{N \to \infty} \frac{\ln P_N(A)}{N} \, . \label{eq:Ia-from-PN}
\end{equation}
The rate function $I(a)$ is convex, with a minimum at $a^* \equiv \langle g(x) \rangle_S = \langle A \rangle/N$. Around the minimal point $a^*$, $I(a)$ usually assumes a parabolic approximation, which matches smoothly with the Gaussian distribution of typical fluctuations \cite{Smith2022}. \new{In general, for large $N$, these properties are independent of the distribution of ${\bm x}_1$ (as long as ${\bm x}_1$ is a continuous random variable).}

One way of calculating the rate function is to use the scaled cumulant generating function (SCGF) $\lambda(k)$, defined as
\begin{equation}
    \left\langle e^{k A} \right\rangle \sim e^{ N \lambda(k)} \, . \label{eq:SCGF}
\end{equation}
The rate function is then found immediately through the Legendre-Fenchel transform \cite{Grassberger1985,Touchette2009,Smith2022,Gutierrez2023,Monthus2024}
\begin{equation}
    I(a) = \sup_k \left[ k a - \lambda(k) \right] \, . \label{eq:legendre-fenchel}
\end{equation}
According to the Gartner-Ellis theorem, the SCGF $\lambda(k)$ is the logarithm of the largest eigenvalue  of the tilted generator of the dynamics $L_k$, defined as \cite{Touchette2009,Laffargue2013,Smith2022,Gutierrez2023,Monthus2024}
\begin{equation}
    L_k \psi_R( \bm{x} ) = \sum_{\bm{z}_i \in f^{-1}(\bm{x})} \frac{e^{k g(\bm{z}_i)} \psi_R(\bm{z}_i)}{|J_f(\bm{z}_i)|} = e^{\lambda(k)} \psi_R( \bm{x} ) \, . \label{eq:tilted-operator}
\end{equation}
Here $\psi_R( \bm{x} )$ is the right eigenfunction of the tilted operator, associated with the eigenvalue $e^{\lambda(k)}$. The tilted operator is not self-adjoint, and therefore the same eigenvalue $\lambda(k)$ is connected to a left eigenfunction, $\psi_L(\bm{x})$, associated with the adjoint tilted operator $L_k^\dagger$ as \cite{Gutierrez2023,Monthus2024}
\begin{equation}
\label{Lkdagger}
    L_k^\dagger \psi_L(\bm{x}) = e^{k g(\bm{x})} \psi_L(f(\bm{x})) = e^{\lambda(k)} \psi_L(\bm{x}) \, .
\end{equation}
The rate functions for $g(x) = x$ have been found in specific cases through a general analytical series expansion \cite{Smith2022}, while in the particular case of the doubling 1d map, $f(x) = 2 x \, \mathrm{mod} \, 1$, the whole spectrum and solutions have been calculated in \cite{Monthus2024}.

\section{Two observables with identical rate functions \label{sec:identical-rate-functions}}

For a practical example, we will consider two standard one-dimensional maps: the fully chaotic logistical map 
\begin{equation}
    f(x) = 4 x \left(1-x\right) \, , \label{eq:def-logistic}
\end{equation} 
and the tent map 
\begin{equation}
    f(x) = 1 - |1 - 2x| \, .     \label{eq:def-tent}
\end{equation}
We will consider the two distinct observables $A_j = \sum_{i=1}^N g_j(x_i)$ where
\begin{equation}
\label{eq:g1g2}
    g_1(x) = \frac{x}{x_c} ~~~ \mathrm{and} ~~~ g_2(x) = \frac{x^2}{x_c^2} \, \, ,
\end{equation}
which are scaled by the non-trivial fixed point $x_c = f(x_c)$ ($x_c \neq 0$). This scaling ensures that, as $N \to \infty$, the domain of $I(a)$ is the same for both observables $a \in [0,1]$, 
with the two extreme cases $a=0$ and $a=1$ corresponding to the sequences $0=x_1=x_2 =\dots$ and $x_c=x_1=x_2 =\dots$, respectively.
For the two specific maps we are considering, $x_c=3/4$ for the logistical map, and $x_c = 2/3$ for the tent map.
We checked that, for the particular choices of $g(x)$ and $f(x)$ studied here, the edges of the domain of $I(a)$ are indeed given by $g(x)$ evaluated at the fixed points of $f(x)$. However, we do not expect this property to hold for all choices of $g(x)$ and $f(x)$.

\new{As a side note, it is well known that the tent and logistic maps are topologically conjugate, meaning there is a transformation $\Phi(x)=\sin^{2}(\pi x/2)$ which maps the chaotic series $x_i$ generated using the tent map onto the series $\tilde{x}_i = \Phi(x_i)$, which is generated by the logistic map \cite{tentMapWiki}.
However, the two cases studied above (for the tent map and logistic map) are genuinely different. This is because, under the transformation $\Phi$, the observable function would also transform,
$g(x) = x/x_c$ in the tent map is transformed into $\tilde{g}(x) =\sin^{2}(\pi x/2)/x_c$ in the logistic map \cite{Smith2022}.}

In order to find the largest eigenvalue of the tilted operator, we must solve the functional eigenvalue Eq.\,(\ref{eq:tilted-operator}). For the logistical map, Eq.\,(\ref{eq:tilted-operator}) becomes
\begin{equation}
     \sum_{y_\pm = \frac{1 \pm \sqrt{1-x}}{2}}\frac{e^{k g\left( y_\pm \right)} \psi_R\left( y_\pm \right)}{4 \sqrt{1-x}} = e^{\lambda(k)} \psi_R(x) \, , \label{eq:tilted-operator-logistical}
\end{equation}
and for the tent map
\begin{equation}
    \frac{e^{k g\left( \frac{x}{2} \right)} \psi_R\left( \frac{x}{2} \right) + e^{k g\left( 1 - \frac{x}{2} \right)} \psi_R\left( 1 - \frac{x}{2} \right)}{2} = e^{\lambda(k)} \psi_R(x) \, . \label{eq:tilted-operator-tent-map}
\end{equation}
Eqs.\,(\ref{eq:tilted-operator-logistical}) and (\ref{eq:tilted-operator-tent-map}) can be solved through an analytical series expansion in powers of $k$ \cite{Smith2022}. This expansion can be performed up to any order of $k$,  and for completeness, in Appendix \ref{appendix:analytical-continuation} we perform the first three orders explicitly for the tent map. Using the series solution, we calculate the rate function using the Legendre-Fenchel transform in Eq.\,(\ref{eq:legendre-fenchel}). 
The solution is obtained as a power series in $a-a^*$. The radius of convergence of this series does not cover the entire domain $a\in [0,1]$.
In order to extend the range of validity to the entire domain $0 < a < 1$, we perform a quasi-analytical continuation using the higher orders (see Appendix \ref{appendix:analytical-continuation} for further details).
An alternative method to calculate the rate function is the power method, which consists of repeatedly applying the tilted operator to an initial density (the IM) until the growth factor stabilizes as the dominant eigenvalue. Specifics of the implementation of the power method are found in Appendix \ref{app:power-method}. 
The two methods described above yield results that exactly coincide.

\begin{figure*}
    \centering
    \includegraphics[width=0.75\linewidth]{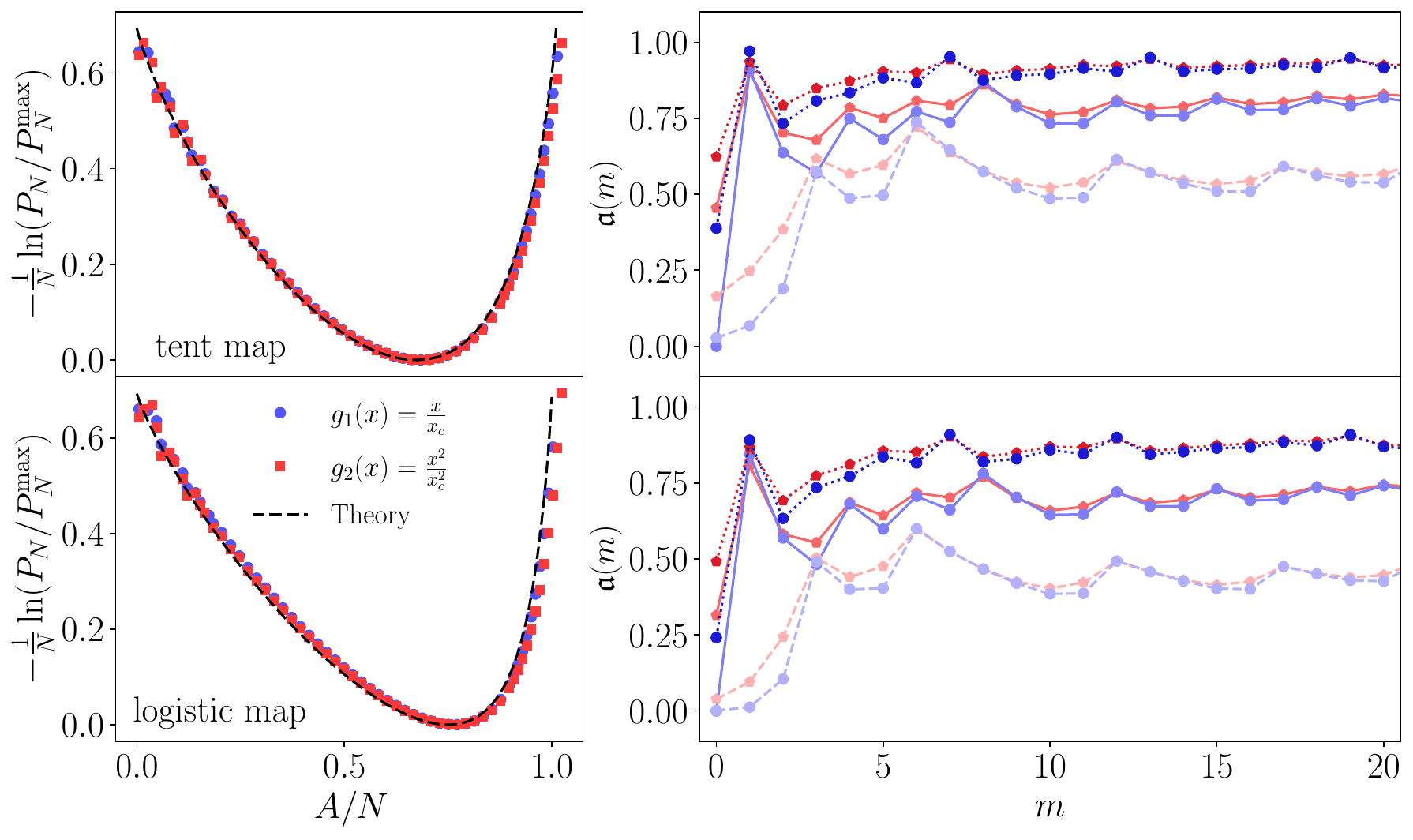}
    \caption{
    Rate function similarity for two distinct observables on the left panels. Symbols show $I_N(a) \equiv -\ln (P_N(A)/P_N^{\mathrm{max}})/N$ obtained from backward–trajectory Monte‑Carlo sampling (the division by the maximum value helps removing the effect of the pre-factors) for the tent map (upper) and the logistic map (bottom) with $N=50$. Blue circles correspond to $g_1(x)=x/x_c$, red squares to $g_2(x)=x^{2}/x_c^{2}$, where $x_c$ is the non‑trivial fixed point of the respective map. The dashed lines are the exact rate functions obtained using Eq.~(\ref{eq:legendre-fenchel}), with $\lambda(k)$ obtained through one of two different analytic methods: The power method and the perturbative $k$ expansion, see Appendix \ref{appendix:analytical-continuation}.
    The collapse of the theoretical and numerical data into a single curve confirms that, remarkably, both observable functions $g_1$ and $g_2$ share the same rate function. The partial sum of $A_i$, $\mathfrak{a}(m)$ (see Eq.\,(\ref{eq:a-cumulative})), related to $g_1(x)$ (blue) and $g_2(x)$ (red) is shown on the right panels for three distinct chaotic sequences, distinguished by line style and tone, lighter to darker representing small to large values of $A_i$. We can see that for large $m$, both averages converge to the same value.}
    \label{fig:dynamical_observable_power}
\end{figure*}

The rate function can also be calculated numerically by obtaining the probability distribution $P_N(A_{i})$, associated with $g_{i}$, directly from the trajectories for fixed values of $N$. We use a Monte-Carlo method \cite{Leitao2014,Leitao2017} to generate these trajectories backward (i.e., beginning from $x_N$), while biasing them toward atypical values of $A$ \cite{Grassberger1985,Smith2022}, see Appendix \ref{appendix:MC} for more details. Using the numerically-computed distributions, we obtain good estimates for the rate functions $I_1(a)$ and $I_2(a)$ through numerical computations of the right-hand side of Eq.\,(\ref{eq:Ia-from-PN}) for finite $N$. To our great surprise, we found a striking result: In each of the two maps, the two rate functions exactly coincide $I_1(a) = I_2(a)$, as seen in the panels of Fig.\,\ref{fig:dynamical_observable_power}.
As a consistency check, one can also check that the formulas \eqref{eq:mean-green-kubo} and \eqref{eq:variance-green-kubo}, for the mean and variance respectively, also give the same result for the two observable functions \eqref{eq:g1g2} for each of the two maps.
It is important to note that the exact distributions $P_N(A_1)$ and $P_N(A_2)$ do not coincide at finite $N$ (this is easy to check explicitly, by considering e.g.~the simple case $N=1$). Therefore, the coincidence is only approximate and occurs at the level of the rate functions, describing the $N \gg 1$ behavior.

This remarkable result may lead one to naively conjecture that the rate function $I(a)$ is independent of the observable function $g(x)$, up to a simple rescaling. However, one can immediately rule out this conjecture by considering a general power observable function $g_\alpha(x) = x^\alpha/x_c^\alpha$. The average \eqref{eq:mean-green-kubo} of this observable is identical for $\alpha=1$ and $\alpha=2$, however, any other value of $\alpha$ has a different mean. This is sufficient to show that the respective rate function is going to be different. We also observe that this property is map dependent.
Indeed, we checked that if $f(x)$ is the doubling map, as we verified numerically, then $I_1(a) \neq I_2(a)$ for the two considered observable functions \eqref{eq:g1g2}. In the next section, we will answer the two key questions: how do two completely distinct functions $g(x)$ lead to observables that follow the same rate function? What is the connection between the two functions and can it be generalized?

\section{The striking similarity and $\Delta g(x)$ 
\label{sec:delta_g}}

Our goal in this Section is to explain the mechanism responsible for the emergence of the striking similarity observed in the previous Section. First, let us denote by $\mathfrak{a}(m)$ the arithmetic mean of the first $m$ terms in the sum that gives $A$, i.e.,
\begin{equation}
    \mathfrak{a}(m) = \frac{1}{m} \sum_{n=1}^m g(\bm{x}_n) \, , \label{eq:a-cumulative}
\end{equation}
where we clearly have that $a = A/N = \mathfrak{a}(N)$.
For the specific examples discussed in the previous Section, if we take a chaotic sequence starting from any initial condition $x_1$, we found that the partial sums for both observable functions \eqref{eq:g1g2} converge to the same value as $m$ becomes larger. This can be seen in the right panel of Fig.~\ref{fig:dynamical_observable_power}. 
This property holds even if $x_1$ is an initial condition that leads to a very unlikely value of the observables $A_j$, and it
hints that the key to understanding this effect lies in the difference between these two observables.
Therefore, we consider the difference between the two observable functions, that is,
\begin{equation}
    \Delta g(x) \equiv g_1(x) - g_2(x) = \frac{x(x_c - x)}{x_c^2} \, . \label{eq:difference-g1-g2}
\end{equation}
We now show that this observable function is of a special form. It turns out that, for both logistic and tent maps, we can rewrite the previous equation as
\begin{equation}
    \Delta g(x) = h(x) - h(f(x)) \, . \label{eq:null-eigenvalue}
\end{equation}
\new{Since Eq.\,(\ref{eq:difference-g1-g2}) is a polynomial, it is possible to find the respective $h(x)$ functions by proposing a general polynomial form $h(x) = a_1 x + a_2 x^2$. Plugging this expression into Eq.\,(\ref{eq:null-eigenvalue}) allows to find that}
\sout{where} for the tent map $h(x) = 3x^2/4 - 3x/2$, and for the logistic map $h(x) =- 4x/9$. 
\new{The task of finding the particular $h(x)$ function for two general observables is not always straightforward.}
We write the difference between the two observables $\Delta A$, as,
\begin{equation}
    \Delta A \equiv A_1 - A_2 = \sum_{n=1}^{N}\Delta g(x_{n}) \, ,
\end{equation}
and the unique form of Eq.\,(\ref{eq:null-eigenvalue}) allows us to write
\begin{align}
    \Delta A &= \left[ h\left(x_{1}\right)-h\left(x_{2}\right) \right] + \left[ h\left(x_{2}\right)-h\left(x_{3}\right) \right] + \dots \nonumber \\
    & \, + \left[ h\left(x_{N}\right)-h\left(x_{N+1}\right) \right] \, ,
\end{align}
where we see that the consecutive terms cancel each other, leaving only the boundary terms, reducing the expression to
\begin{equation}
    \Delta A = h(x_1) - h(x_{N+1}) \, .
\end{equation}
Given that $h(x)$ is a well-behaved function, i.e., does not diverge in $x \in [0,1]$, $\Delta A$ is of order $O(1)$ in the large-$N$ limit. This feature is what leads to the striking similarity we observed, despite both observables being clearly distinct, the relative difference between them is relatively small as $N$ goes to infinity.
In fact, not only do the rate functions coincide, but the realizations of the dynamics that yield a given value of $A_i$ are nearly (at $N \gg 1$) the same for both observables. This can be useful in the context of generating biased simulations, i.e., realizations of the dynamics conditioned on a given atypical value of the observable $A$ \cite{Gutierrez2023}.
For a discussion of the case in which $h(x)$ is not bounded, see Section~\ref{sec:FTLE} below.

Our findings, motivated by specific examples in $d=1$, can be generalized to arbitrary chaotic maps $f$ and (bounded) functions $h$. Indeed, consider a $d$-dimensional map $f(\bm{x})$ and two observable functions $g_1(\bm{x})$ and $g_2(\bm{x})$ that are distinct and whose difference can be written in the form
\begin{equation}
    g_1(\bm{x}) - g_2(\bm{x}) = h(\bm{x}) - h(f(\bm{x})) \, ,
\end{equation}
where $h(\bm{x})$ is a well-behaved function. 
Then, the same argument given above for the particular examples considered there holds and implies that the rate functions for the two observables coincide exactly.

The coincidence can also be understood within the theoretical framework that we described above for calculating the rate function \cite{Smith2022}. This approach yields additional insight, as we now show.
The eigenvalue spectra 
of each observable can be found through their respective tilted operators, see Eq.\,(\ref{eq:tilted-operator}). Each eigenvalue 
is associated with right and left eigenfunctions $\psi_{R}^{(i)}(\bm{x})$ and $\psi_{L}^{(i)}(\bm{x})$ respectively.
As one can check directly from the form of the tilted operator \eqref{eq:tilted-operator}, the right eigenfunctions for the two observables are related by
\begin{equation}
\label{eq:relationRight}
\psi_{R}^{(1)}(\bm{x})=e^{-kh(\bm{x})}\psi_{R}^{(2)}(\bm{x})\,,
\end{equation}
and similarly, the left eigenfunctions are related by
\begin{equation}
\label{eq:relationLeft}
\psi_{L}^{(1)}(\bm{x})=e^{kh(\bm{x})}\psi_{L}^{(2)}(\bm{x})\,.
\end{equation}
As a result, the spectra for the two observables are the same. In particular, the largest eigenvalues $\lambda_i(k)$ are the same and therefore so are the associated rate functions $I_i(a)$. 
\new{As mentioned above, it is easy to verify that these Eqs.\,(\ref{eq:relationRight}) and (\ref{eq:relationLeft}) are correct. However, we provide a derivation of them in Appendix \ref{app:eigenvectorRelation}.}

It is worth noting that the product $\psi_{R}^{(i)}(\bm{x})\psi_{L}^{(i)}(\bm{x})$ of the left and right eigenfunctions is the same for both observables. This implies that the distribution $p\left(\bm{x}_{j}|A_{i}\right)$ of intermediate elements in the sequence $\bm{x}_j$ (with $j \gg 1$ and $N-j \gg 1$) conditioned on either one of the two observables taking a given value $A_i = aN$ [where $k = I'(a)$] is the same for the two observables \cite{Burenev2025a}. The latter result is a consequence of the more general property described above concerning the coincidence of trajectories conditioned on $A_i$.

Importantly, if $h(\bm{x})$ is not bounded, then the relations \eqref{eq:relationRight} and \eqref{eq:relationLeft} between the eigenfunctions may relate between normalizable and nonnormalizable eigenfunctions. As a result, in this case, the two spectra do not necessarily coincide for some $k$'s (and similarly for the SCGFs and rate functions), see Section~\ref{sec:FTLE} below.

\section{The case where $g(\textbf{\textit{x}})$  is a derived function \label{sec:g-derived-function}}

An interesting case to study is when $g(\bm{x})$ itself is a derived function, i.e., it can be written in the form
\begin{equation}
    g^*(\bm{x}) = h(\bm{x}) - h(f(\bm{x}))\, .\label{eq:definition-derived-function}
\end{equation}
We will refer to the corresponding observables $A^* = \sum_{i=1}^N g^*(\bm{x})$ as derived observables.
Following the analysis of the previous section, one can relate $A^*$ with the null observable [corresponding to $g_2(x)=0$]. 
It follows from this analysis that, provided $h(\bm{x})$ is bounded, the rate function for $A^*$ is trivial, and indeed this is not surprising since $A^* = h(\bm{x}_1) - h(\bm{x}_{N+1})$ and it is therefore bounded by two $N$-independent values (as a result $a^* = A^* / N$ cannot take any nonzero, $O(1)$ value in the large-$N$ limit). \new{It is important to note that the choice of initial conditions will have a strong effect on the distribution of $A^*$. This is in contrast to the case where the observable is not derived, where the initial condition is not important in the large-$N$ limit.}

In this section, we first characterize derived functions and then analyze the distributions of associated observables $A^*$, while in the next section, we demonstrate interesting effects that may occur for $h(\bm{x})$ which is not bounded by focusing on the particular case of the FTLE for the logistic map.

\subsection{Characterizing derived functions}

Although we have defined the form of a derived function above, in practice it can be highly nontrivial to determine whether a given function is of this form. It also may be difficult, given a derived function $g^*(\bm{x})$, to find the corresponding $h(\bm{x})$.
It is therefore useful to give alternative characterizations and to study some basic properties of derived functions. This has been done in fairly general settings in the mathematical literature \cite{Livshits1971}, but for completeness we will state and derive the results that will be most useful to us here.

1) If $g^*(\bm{x})$ is derived, then its ensemble-average value on the IM vanishes. This may be seen by a direct calculation,
\begin{align}
\left\langle g^{*}(\bm{x})\right\rangle_S &=\int d\bm{x}g^{*}(\bm{x})p_{s}(\bm{x}) \nonumber\\
&=\int d\bm{x}h(\bm{x})p_{s}(\bm{x})-\int d\bm{x}h\left(f\left(\bm{x}\right)\right)p_{s}(\bm{x}) \, .
\end{align}
By changing the integration variable in the second integral on the right-hand side to $f\left(\bm{x}\right)$, it is straightforward to show that the result coincides with the first integral.
As a result, the mean of $A^*$ vanishes exactly.

2) The variance of $A^*$ grows sublinearly with $N$. This may be seen through the Green-Kubo formula \eqref{eq:variance-green-kubo}. Indeed, for a derived observable function, one has
\begin{eqnarray}
    \langle g^*(\bm{x}_{0})g^*(\bm{x}_{\delta})\rangle_{S}&=&\langle h(\bm{x}_{0})h(\bm{x}_{\delta})\rangle_{S}-\langle h(\bm{x}_{0})h(\bm{x}_{\delta+1})\rangle_{S}\nonumber\\
    &-&\langle h(\bm{x}_{1})h(\bm{x}_{\delta})\rangle_{S}+\langle h(\bm{x}_{1})h(\bm{x}_{\delta+1})\rangle_{S}\nonumber\\
    &=&2c_{\delta}-c_{\delta-1}-c_{\delta+1} 
\end{eqnarray}
where
\begin{equation}
    c_{\delta}=\langle h(\bm{x}_{0})h(\bm{x}_{\delta})\rangle_{S} \, ,
\end{equation}
and as a result, the sum in \eqref{eq:variance-green-kubo} vanishes:
\begin{equation}
\label{eq:GreenKuboZero}
    \sum_{\delta=-\infty}^{\infty}\left\langle g^{*}\left(\bm{x}_{0}\right)g^{*}\left(\bm{x}_{\delta}\right)\right\rangle _{S}=\sum_{\delta=-\infty}^{\infty}\left(2c_{\delta}-c_{\delta-1}-c_{\delta+1}\right)=0
\end{equation}
where we used the fact that $\left\langle g^{*}(\bm{x})\right\rangle_S = 0$ and that $\bm{x}_0$ and $\bm{x}_\delta$ become uncorrelated for large values of $\delta$.

Properties 1) and 2) clearly hold if $h$ is bounded. However, they are expected to hold also if 
$h$ is unbounded, provided all integrals and sums in the equations above converge.

3) The average of $g^*(\bm{x})$ over any periodic orbit vanishes. Indeed, if
$\bm{x}_{1}, \dots, \bm{x}_{p}$ is a periodic orbit of period $p$ (so $f(\bm{x}_{p}) = \bm{x}_{1}$) then
\begin{equation}
    \label{eq:periodic}
    \sum_{i=1}^{p}g^*\left(\bm{x}_{i}\right)=\sum_{i=1}^{p}h\left(\bm{x}_{i}\right)-\sum_{i=1}^{p}h\left(f\left(\bm{x}_{i}\right)\right)=0 \, .
\end{equation}

In the converse direction, if $g^*$ satisfies the third property (for all periodic orbits), then, under certain regularity conditions on the functions $g^*$ and $f$, it follows from Livshits's theorem \cite{Livshits1971,Livshits1972} that $g^*$ is a derived function.
It would be interesting to check if similar converse statements hold regarding the other two properties.

We mention now that one could perhaps try to generalize the concept of a derived function to functions of the type
$h(\bm{x}) - h(f(f(\bm{x}))$ etc.
However, such functions are in fact also derived (according to the definition given above),
\begin{equation}
    g^{*}(\bm{x})=\tilde{h}(\bm{x})-\tilde{h}\left(f\left(\bm{x}\right)\right)
\end{equation}
with
\begin{equation}
    \tilde{h}(\bm{x})=h(\bm{x})+h\left(f\left(\bm{x}\right)\right) \, .
\end{equation}
This argument may straightforwardly be extended to any function that is given by a linear combination
\begin{equation}
    \sum_{i=1}^{M}\alpha_{i}h\left(\bm{x}_{i}\right)
\end{equation}
such that the sum of the coefficients vanishes, that is, $\sum_{i=1}^{M}\alpha_{i}=0$.

\subsection{Distribution of typical fluctuations}

Let us now analyze the distribution $P_N(A^*)$  of derived observables $A^*$. Since $A^* = h(\bm{x}_1) - h(\bm{x}_{n+1})$, for any function $h(\bm{x})$,
we can approximate $P_N(A^*)$ at large $N$ by considering the point $\bm{y} \equiv \bm{x}_{N+1}$ and the initial point $\bm{x} \equiv \bm{x}_1$ to be statistically independent.
Using that each of the two points is distributed according to the IM, we then obtain
\begin{equation}
    P_N(A^*) \simeq \int  p_S(\bm{x}) p_S(\bm{y}) \delta \left[A^* - h(\bm{x}) + h(\bm{y}) \right] d\bm{x} d\bm{y} \, . \label{eq:PNA}
\end{equation}
This simply means that the distribution is given by the convolution
\begin{equation}
P_N\left(A^*\right) \simeq \int p_{h}\left(u\right)p_{h}\left(u-A^*\right)du
\end{equation}
where
\begin{equation}
p_{h}\left(u\right)=\int\delta\left(u-h\left(\bm{x}\right)\right)p_{S}\left(\bm{x}\right)d\bm{x}   
\end{equation}
is the distribution of $u = h(\bm{x}_i)$. The approximation \eqref{eq:PNA} implies several interesting general properties of $P_N(A^*)$. First of all, $P_N(A^*)$ becomes independent of $N$ at $N \to \infty$. Furthermore, $P_N(A^*)$ is mirror symmetric $P_N(A^*) \simeq P_N(-A^*)$.
It follows that the mean of the distribution is zero (as was already shown above), but also that the variance is twice the variance of $h(\bm{x}_1)$, 
\begin{equation}
\text{Var}(A^*) \simeq \text{Var}(h(\bm{x}_1)) + \text{Var}(h(\bm{x}_{N+1})) = 2\text{Var}(h(\bm{x}_1) \, .
\end{equation}
In particular, the variance is independent of $N$ (at $N \gg 1$) which is consistent with the vanishing result of the Green-Kubo formula, Eq.~\eqref{eq:GreenKuboZero}.

In Sec.~\ref{sec:identical-rate-functions}, for the 1d tent and logistic maps, we have used the first and second powers of $x$ as observable functions as seen in Eq.~(\ref{eq:g1g2}). Following from those examples, let us now analyze the distribution of the observable generated by the derived function $g^*(x)$ which is given by the difference between the two observable functions from Eq.~(\ref{eq:g1g2}), see Eq.~(\ref{eq:difference-g1-g2}). The full expression is
\begin{equation}
    g^*(x) = \frac{x}{x_c} - \frac{x^2}{x_c^2} = h(x) - h(f(x)) \, , \label{eq:1d-derived-function}
\end{equation}
where the functions $h(x)$ were found above and given just after Eq.~\eqref{eq:null-eigenvalue} and, for convenience, are given here as well.

For the tent map, which is defined in Eq.\,(\ref{eq:def-tent}), we find that
\begin{equation}
    h(x) = \frac{3}{4}x^2 - \frac{3}{2}x \, . \label{eq:h(x)-tent}
\end{equation}
Eq.\,(\ref{eq:PNA}) relies on the assumption that the correlation between starting point $x_1$ and final $x_{N+1}$ is negligible. We can show this directly by calculating
\begin{equation}
    \langle h(x) \rangle_S = - \frac{1}{2} \, \, \, \mathrm{and} \, \, \, \langle h(x_1) h(x_n) \rangle_S = \frac{2^{-(2n+1)}}{10} + \frac{1}{4} \, ,
\end{equation}
which leads to the correlation
\begin{equation}
    \langle h(x_1) h(x_n) \rangle_S - \langle h(x) \rangle_S^2 = \frac{2^{-(2n+1)}}{10} \, ,
\end{equation}
which shows an exponential decay. Therefore, even for moderately-large values of $N$, we expect Eq.\,(\ref{eq:PNA}) to hold. Replacing $h(x)$ defined in Eq.\,(\ref{eq:h(x)-tent}) into Eq.\,(\ref{eq:PNA}) and performing the integral, we find
\begin{equation}
    P_N(A^*) \simeq \frac{1}{3} \ln \left[ \frac{3 + \sqrt{9 - 12 |A^*|} - 2 |A^*|}{2 |A^*|} \right] \,  \label{eq:PNA_tent} .
\end{equation}

\begin{figure*}
    \centering
    \includegraphics[width=0.75\linewidth]{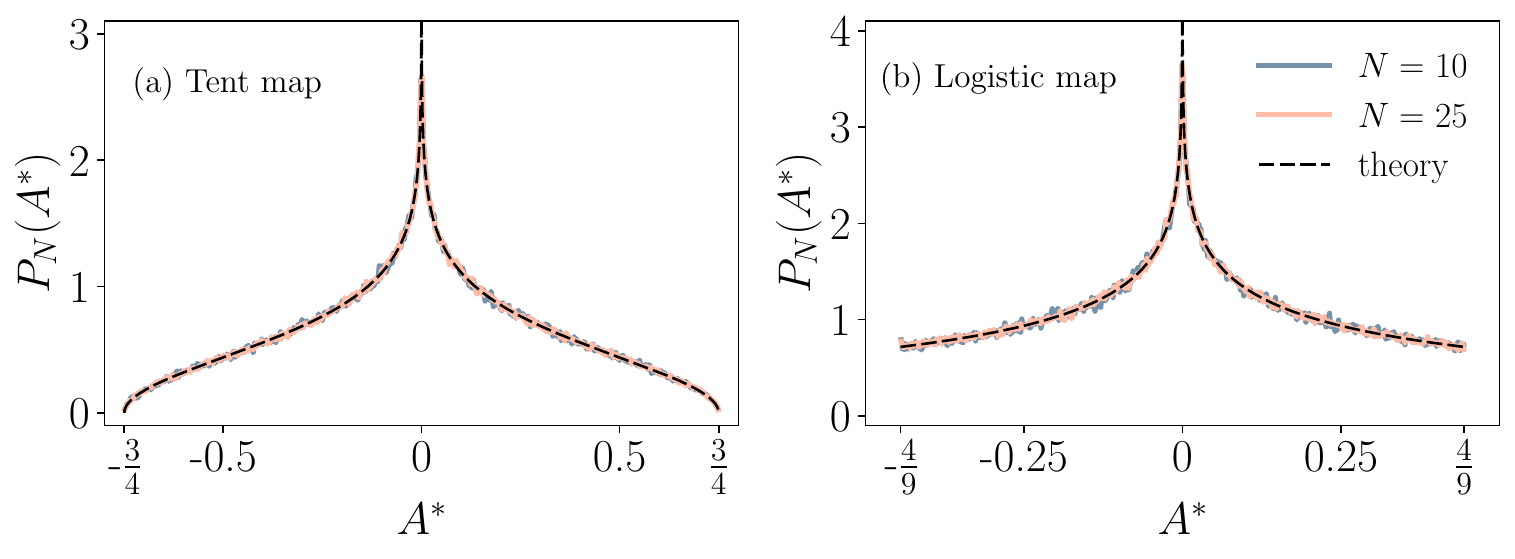}
    
    \caption{Probability density of the derived observable $A^*$, corresponding to the observable function \eqref{eq:1d-derived-function}, for (a) the tent map and (b) the logistic map.  
Solid lines are numerical results for $N=10$ (dark color) and $N=25$ (light color), the solid black line is the parameter‑free prediction obtained from Eq.~(\ref{eq:PNA}) with the explicit forms in Eqs.~(\ref{eq:PNA_tent}) and (\ref{eq:PNA_logistic}). The caption for panel (b) applies to both figures.
Because $A^*=h(x_1)-h(x_{N+1})$ with bounded $h$, the distribution becomes independent of $N$ at large $N$ and therefore mirror–symmetric around zero. We can also see clearly that $A^*$ is bound between $\pm \left[\max_{x}\left(h\left(x\right)\right)-\min_{x}\left(h\left(x\right)\right)\right]$ for both maps.}
    \label{fig:PNA}
\end{figure*}

For the logistic map, the function $h(x)$ in Eq.\,(\ref{eq:1d-derived-function}) is
\begin{equation}
    h(x) = - \frac{4}{9} x \, , \label{eq:h(x)-logistic}
\end{equation}
and the correlation function vanishes identically,
\begin{equation}
    \langle h(x_1) h(x_n) \rangle_S = 0 \, .
\end{equation}
Plugging Eq.\,(\ref{eq:h(x)-logistic}) into Eq.\,(\ref{eq:PNA}), we find
\begin{equation}
    P_N(A^*) \simeq \frac{9}{2 \pi^2} K\left(1 - \frac{81 {A^*}^2}{16} \right) \, , \label{eq:PNA_logistic}
\end{equation}
where $K(x)$ is the complete elliptic integral of the first kind \cite{EllipticKWolfram}. We show the validity of Eqs.\,(\ref{eq:PNA_tent}) and (\ref{eq:PNA_logistic}) in panels (a) and (b), respectively, of Fig.\,\ref{fig:PNA}, where these predictions are compared to numerical computations of $P_N(A^*)$ for $N=10,25$.
In both cases, the predicted $P_N(A^*)$ diverges at $A^* = 0$. However, this property does not necessarily hold in general; it depends on if $p_h(u)$ diverges near the edges of its support, and if so, how strong this divergence is.

The approximation of $x_1$ and $x_{N+1}$ as statistically independent may break down in the large-deviations regime of $A^*$. This can happen because unusual values of $A^*$ are associated with unusual values of $x_1$ for which the statistical-independence approximation is incorrect.
However, for the two examples given above, $h(x)$ is bounded and as a result, $A^*$ is bounded and therefore the rate function is trivial (and could be considered to be $I(a)=\infty$ for all nonzero $a$).
Eq.~\eqref{eq:PNA} therefore gives a complete description of the distribution.
If $h(x)$ is not bounded, Eq.~\eqref{eq:PNA} still describes the typical fluctuations of $A^*$ correctly, but the large-deviations regime may be nontrivial, as demonstrated below in Section \ref{sec:FTLE}.

We conclude this subsection by an analysis of the eigenfunctions of the tilted Frobenius-Perron operator, for derived observables with a bounded $h(\bm{x})$.
Using the results of Sec.\,\ref{sec:delta_g} with $g_1(\bm{x}) = g^*(\bm{x})$ and $g_2(x)=0$, we find that the SCGF is trivial $\lambda(k)=0$ for all derived functions $g^*(x)$, provided that $h(\bm{x})$ is bounded. 
For the observable function $g_2(\bm{x}) = 0$, the tilted Frobenius-Perron operator trivially coincides with the untilted one. As a result, the corresponding right and left eigenfunctions associated with the largest eigenvalue $e^{\lambda(k)} = 1$, for all values of $k$ are the IM and the constant function, respectively. Therefore, from Eqs.\,(\ref{eq:relationRight}) and (\ref{eq:relationLeft}), the right and left eigenfunctions for the derived observable $g_1(\bm{x})$ are given by
\begin{equation}
    \psi_R(\bm{x}) = e^{- k h(\bm{x})} p_S(\bm{x}) \, \, \, \mathrm{and} \, \, \, \psi_L(\bm{x}) = e^{k h(\bm{x})} \, , \label{eq:derived-eigenfunctions}
\end{equation}
\new{where we used the well-known fact that the left eigenfunction for the Frobenius-Perron operator is a constant.}
The validity of the previous equation can be immediately verified by plugging the expressions into Eq.\,(\ref{eq:tilted-operator}).

{
\section{Application to random walks generated by an open chaotic map \label{sec:diffusion-chaos}}

It is usually the case that random walks generated by open chaotic maps (i.e., maps whose image lies outside the interval on which they are defined) display diffusive or anomalous diffusive behavior at long times \cite{Klages2010_Chaos2Diffusion}.
As a simple example, let us consider a family of piecewise linear maps $f(x)$, which we initially define for $x \in [0,1]$ through
\begin{equation}
    f\left(x\right)=\begin{cases}
-bx, & 0<x<x_{*},\\[1mm]
\delta - bx, & x_{*}<x<1.
\end{cases} \label{eq:diffusive-map-general-without-mod}
\end{equation}
$f(x)$ may lie outside the interval $[0,1]$, so we extend its definition to the entire real line through \cite{Klages2010_Chaos2Diffusion}
\begin{equation}
    f(x + n) = f(x) + n
\end{equation}
for integer $n$ and $x \in [0,1]$.
Qualitatively similar piecewise linear maps have been studied extensively, see \cite{Fujisaka1982a,Klages1995,Klages1999}.
Under certain assumptions (in particular, $|b| > 1$) $f(x)$ is expected to be chaotic and ergodic.

We interpret $x_i$ as the position of the random walker (RW) at time $i$. Whenever the RW exits an interval $[n,n+1]$ (for integer $n$) we will say that it has ``hopped''. The direction of the hop is right (left) if $x_i \in [n,n+1]$ and $f(x_i) \in [n+1,n+2]$ ($f(x_i) \in [n-1,n]$) (in what follows, we assume that the parameters of $f$ are such that only single hops are possible).
Fundamental observables of interest for such random walks are the effective diffusion coefficient which describes the long-time behavior of the position of the RW, the ``escape rate'' that describes the decay of the probability for no hops to occur for many consecutive steps, and the ``dynamical activity'' which describes the total number of hops performed \cite{Perez-Espigares2019,Casert2021,Monthus2022,Monthus2024a}.

In what follows, we will specialize to the particular choice of parameters $b=(1+\sqrt{5})/2 = 1.618\dots$ is the golden ratio $\golden$, $\delta = 2$ and $x_\star = 2- b = 0.372\dots$, see Fig.~\ref{fig:diff-model}. For these parameters we were able to obtain analytical results and found surprising connections to the theoretical framework of derived observables that we developed above.}

{
A mathematically equivalent formulation of the problem is to consider a sequence $\tilde{x}_1, \dots, \tilde{x}_N$, where $\tilde{x}_i$ is the fractional part of $x_i$. This sequence is generated by the map $\tilde{f}(\tilde{x}) = f(\tilde{x}) \mod 1$ (in what follows, we suppress the tildes on the $x_i$'s to lighten the notation). This map is shown in red in panel (a) of Fig.~\ref{fig:diff-model}. The total number of hops to the left or right then take the form of a dynamical observable \eqref{Adef}, as do the activity and the integer part of the RW's position.
It is convenient to write the associated observable functions $g(x)$ using the indicator function for the interval $[0,x_*]$
\begin{equation}
    H(x) = \begin{cases}
        1 \, , & x < x_\star\, , \\[1mm]
        0\, , & x > x_\star\, .
    \end{cases} \label{eq:indicator-function}
\end{equation}
Indeed, the observable function for the left hops is $g_\mathrm{left}(x) = H(x)$. For right hops, the observable function $g_\mathrm{right}(x)$ is the indicator function for the interval $[x_*,1-x_*]$, and one can check that it can also be written as $g_\mathrm{right}(x) = H(f(x))$, see panel (b) of Fig.~\ref{fig:diff-model}. The observable function for the activity is $g_\mathrm{left}(x) + g_\mathrm{right}(x)$. For future convenience, we will define a scaled activity as $g_\mathrm{activity}(x) = (g_\mathrm{left}(x) + g_\mathrm{left}(x))/2$.

\begin{figure}
    \centering
    \includegraphics[width=\linewidth]{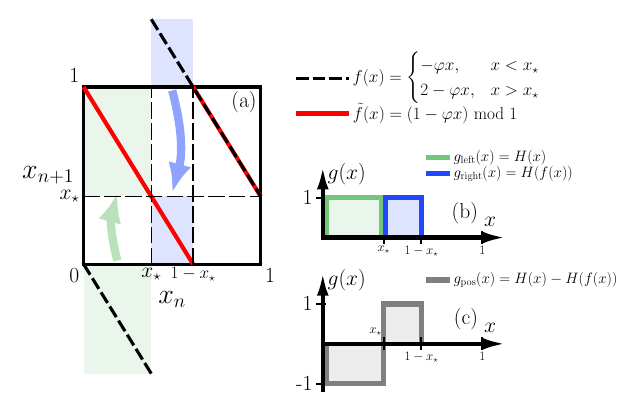}
    \caption{Panel (a): the open chaotic map in Eq.\,(\ref{eq:diffusive-map-general-without-mod}) for $\delta=2$ and $b = (1+\sqrt{5})/2$ (which is the golden ratio $\golden$) in dashed black lines. The folded map $\tilde{f}(x) = f(x) \mod 1$, which governs the dynamics bound to the unit interval, is shown with a solid red line. Panel (b): the observable functions that keep track of hops to the left $g_\mathrm{left}(x)$ (light green) and to the right $g_\mathrm{right}(x)$ (dark blue). Both can be represented using the indicator function $H(x)$ defined in Eq.\,(\ref{eq:indicator-function}). Panel (c): the actual position $g_\mathrm{pos}(x) = g_\mathrm{right}(x) - g_\mathrm{left}(x)$, represented by the difference between right and left hops. This observable function is clearly derived, as seen from Eq,\,(\ref{eq:definition-derived-function}).} 
    \label{fig:diff-model}
\end{figure}

It is easy to see that the difference between any two of these three observable functions is a derived function:
\begin{eqnarray}
\label{gleftminusgright}
g_\mathrm{left}(x) - g_\mathrm{right}(x) &=& H(x) - H(f(x)) \, ,\\
g_\mathrm{activity}(x) - g_\mathrm{right}(x) &=& g_\mathrm{left}(x) - g_\mathrm{activity}(x) \nonumber\\
&=& (H(x) - H(f(x)) )/2
\end{eqnarray}
all of which are compatible with Eq.~(\ref{eq:definition-derived-function}).
Therefore, the full distribution, $P_N(A_j),$ of the corresponding integrated observables, $A_j = \sum_i g_j(x_i)$, will be described by the same large deviation rate function, a fact that is confirmed numerically in panel (a) of Fig.\,\ref{fig:diffusion-rate-functions}.
To find the rate function for all three cases, we only need to solve the problem for one. We will now focus on $g_\mathrm{right}(x) = H(f(x))$.

The IM $P_s(x)$ for the $\tilde{f}(x)$ map can be found through the FPO, defined explicitly as,
\begin{equation}
    \hat{\mathcal{L}}_\mathrm{FP} P_s(x) = \begin{cases}
        \frac{P_s(\frac{1-x}{\golden})}{\golden} \, , & x < x_\star \, , \\[2mm]
        \frac{P_s(\frac{1-x}{\golden})}{\golden} + \frac{P_s(\frac{1-x}{\golden} + 1 - x_\star)}{\golden}\, ,  & x > x_\star \, .
    \end{cases} 
\end{equation}
To find the IM (up to an overall normalization constant), we propose an ansatz where $P_s(x) \propto 1$ for $x<x_\star$ and $P_s(x) \propto c_1$ for $x>x_\star$, and plug this ansatz into the FPO equation, we find that $c_1 = 2 - x_\star$. Therefore, the IM, explicitly, is
\begin{equation}
    P_s(x) \propto \begin{cases}
        1 \, , & x < x_\star \, ,\\[1mm]
        2 - x_\star\, , & x > x_\star\, .
    \end{cases} \label{eq:IM-golden-ratio}
\end{equation}
The tilted operator for $g_\mathrm{right}(x)$, following Eq.~(\ref{eq:tilted-operator}), is
\begin{equation}
    \hat{\mathcal{L}}_k \psi_k(x) = e^{k H(x)} \hat{\mathcal{L}}_\mathrm{FP} \, .
\end{equation}
Since the $e^{k H(x)}$ is also defined piece-wise with respect to $x_\star$, the eigenfunctions can be found by proposing a similar ansatz as the one we used for the IM. The SCGF is found to be
\begin{equation}
    \lambda(k) = \ln \left\{ \frac{1}{4} \left( \sqrt{5} - 1 \right) \left(1 + \sqrt{1 + 4 e^{k}} \right) \right\} \, , \label{eq:eigenvalue-diffusion}
\end{equation}
which, as we see in panel (b) of Fig.~\ref{fig:diffusion-rate-functions}, agrees with the numerical results. The associated eigenfunctions are
\begin{equation}
\label{psiksolRW}
    \psi_k(x) \propto \begin{cases}
        1\, , & x < x_\star \, ,\\[1mm]
        e^{\lambda(k) - k} (2 - x_\star)\, , & x > x_\star\, .
    \end{cases} 
\end{equation}
The rate function $I(a)$ is obtained by applying the Legendre-Fenchel transform (\ref{eq:legendre-fenchel}) to the SCGF (\ref{eq:eigenvalue-diffusion}), see Appendix \ref{app:Legendre}:
\begin{eqnarray}
\label{IofaRW}
I\left(a\right)&=&\ln\left(\frac{1}{a-1}+2\right)-2a\ln(1-2a) \nonumber\\
&+&a\ln\left(a(1-a)\right)+\ln\golden \, .
\end{eqnarray}
In panel (a) of Fig.\,\ref{fig:diffusion-rate-functions}, we see that the numerically-computed PDFs for $N=40$ show good agreement with the analytical rate function for all three observables defined above.

The rate function for the activity describes, in particular, the exponential decay rate for the probability for no hops to occur -- the escape rate -- which is given by $I(0)$. The latter is given by
\begin{equation}
    I(0) = \ln \golden = 0.4812\dots \,,
\end{equation}
coinciding with the Lyapunov exponent of $\tilde{f}$. Physically, this corresponds to an initial condition $\tilde{x}_1$ that is exponentially close to the fixed point $4/(3+\sqrt{5})$ of $\tilde{f}$.
One can similarly show that $I(1/2) = \ln \golden$, which corresponds to an initial condition that is exponentially close to the other fixed point, $2/(3+\sqrt{5})$.

\begin{figure}
    \centering
    \includegraphics[width=\linewidth]{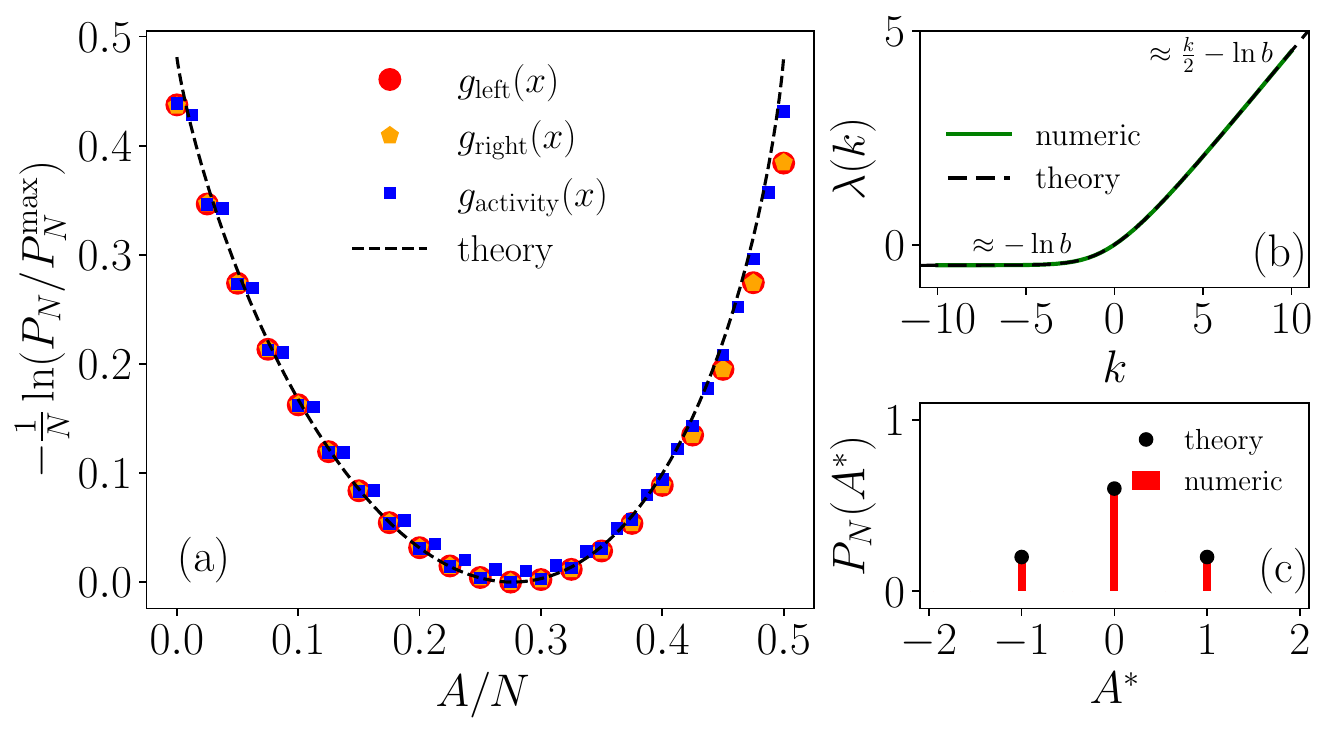}
    \caption{Rate function similarity is shown for the three distinct cases of $g_\mathrm{right}(x)$, $g_\mathrm{left}(x)$, and $g_\mathrm{activity}(x)$ on panel (a). The numerical PDFs $P_N(A)$, colored markers, are obtained through forward-trajectory sampling, with initial position sampled from the IM in Eq.\,(\ref{eq:IM-golden-ratio}), for $N=40$ with $10^{10}$ samples. The theoretical rate function, dashed black line, is obtained by calculating the Legendre-Fenchel transform (\ref{eq:legendre-fenchel}) of the SCGF (\ref{eq:eigenvalue-diffusion}), which is also shown in panel (b). The derived position observable is shown in panel (c), where there are only three possible values of $A^* = 0, \pm1$. We can see clearly that the theoretical prediction (black markers) in Eq.\,(\ref{eq:derived-prob-diffusion-pos}) matches our numerical findings (red bars).}
    \label{fig:diffusion-rate-functions}
\end{figure}

The integer part of the RW's position, $A^*$, may be described by the observable function $g_\mathrm{pos}(x) = g_\mathrm{right}(x) - g_\mathrm{left}(x)$, which equals +1 (-1) when the particle hops to the right (left), as shown in panel (c) of Fig.~\ref{fig:diff-model}. From Eq.~\eqref{gleftminusgright}, we can see that this is a derived observable, and thus the distribution of $A^*$ can be found using Eq.~(\ref{eq:PNA}) [with the IM \eqref{eq:IM-golden-ratio}], as
\begin{equation}
    P_N(A^*) = \begin{cases}
        \frac{x_\star}{1 + (1-x_\star)^2} \, , & A^* = \pm 1 \, , \\[2mm]
        \frac{(2-x_\star)(1-x_\star)}{1 + (1-x_\star)^2} \, ,& A^* = 0 \, .
    \end{cases} \label{eq:derived-prob-diffusion-pos}
\end{equation}
The physical meaning of the previous equation is that a jump to the right in the $i$-th iteration occurs if and only if a jump to the left occurs in the $i+1$-th iteration.
Therefore, the position of the random walker will always remain at $0$ or $\pm 1$, regardless of the number of steps $N$. In particular, the effective diffusion coefficient vanishes. 

We emphasize that, for generic choices of parameters in the definition \eqref{eq:diffusive-map-general-without-mod} of the map, none of the observables considered here nor the differences between them would be derived, and as a result, the RW is in general diffusive at long times. The qualitative phenomena observed here (which essentially boil down to the fact that $g_\mathrm{pos}(x)$ is a derived function) hold more generally provided
\begin{equation}
b = 1 / (1-x_*), \quad \delta = 1+x_* + b x_*
\end{equation}
for any $0 < x_* \le 2-\golden$. However, for the particular choice
$x_* = 2-\golden$, the problem is analytically solvable due to the especially simple form of the eigenfunctions \eqref{psiksolRW}.

}

\section{The finite time Lyapunov exponent (FTLE) in the logistic map \label{sec:FTLE}}

In this Section, we shall study the finite-time Lyapunov exponent (FTLE) in the logistic map. The Lyapunov exponent
\begin{equation}
\ell = \lim_{N \to \infty} \frac{1}{N} \sum_{n=1}^N \ln \sigma(\bm{x}_n) \, , \label{eq:actual-lyapunov}
\end{equation}
\new{where $\sigma(\bm{x})$ is the largest singular value of the Jacobian of the map $f(\bm{x})$}, is an important quantity that measures the system's sensitivity to initial conditions.
Assuming ergodicity, the Lyapunov exponent is independent of the choice of $x_1$, and it is given by the ensemble average
\begin{equation}
\ell=\left\langle \ln \sigma (\bm{x}_{1})\right\rangle _{S} \, .
\end{equation}
For a finite $N$, the FTLE is defined explicitly as
\begin{equation}
    \ell_N(\bm{x}_1) = \frac{1}{N} \sum_{n=1}^N \ln \sigma(\bm{x}_n) \, . \label{eq:def-FTLE-true}
\end{equation}
In general, the FTLE depends on the initial condition, and it is natural to study the distribution of the FTLE, assuming, e.g., that $\bm{x}_1$ is sampled from the IM. Statistical properties of the FTLE have been extensively studied in the past decades \cite{Eckmann1986,Haller2001,Bec2006,Laffargue2013,Biferale2014,Johnson2015,Pazo2016,DeCruz2018,Yoshida2021}, and in particular, for the logistic map \cite{Grassberger1985,Prasad1999,Anteneodo2004,Lai2007,Gutierrez2023}.
At $N \to \infty$, the FTLE converges to the Lyapunov exponent of the map. However, at large but finite $N$, one can study the fluctuations of the FTLE around its mean value. For the logistic map, this was done in \cite{Anteneodo2004} by taking the large-$N$ limit of an exact solution. Remarkably, it was found that the variance of the FTLE scales anomalously with $N$ at $N\gg1$. Moreover, at $N \gg 1$ the distribution of the FTLE was found to become mirror symmetric around its mean \cite{Anteneodo2004}. Large deviations of the FTLE also display interesting properties, including singularities in the rate function and SCGF, which may be interpreted as a dynamical phase transition \cite{Gutierrez2023}. In this Section, we build on the results of the previous sections to shed light on these special properties of fluctuations of the FTLE in the logistic map.

The FTLE, as defined in Eq.\,(\ref{eq:def-FTLE-true}), is not a derived observable. However, it turns out that in the specific case of the logistic map \eqref{eq:def-logistic}, we can define a new observable using the function $h(x) = - \ln f(x)$, which yields the observable function
\begin{equation}
\label{gStarDefFTLE}
   \new{ g^*(x) =  h(x)  - h(f(x)) = 2 \ln |f'(x)| - \ln 4 \, .}
\end{equation}
The associated observable is a scaled and shifted version of the FTLE:
\begin{equation}
    A^* = \sum_{i=1}^N g^*(x) = 2 N \, \ell_N - N \ln 4 \, . \label{eq:scaled-FTLE}
\end{equation}
Here, unlike what was studied in the previous section, $h(x)$ is not bounded. This fact leads to some nontrivial consequences, as shown below. For the logistic map, using Eq.\,(\ref{eq:actual-lyapunov}), we find that the Lyapunov exponent is $\ell = \langle \ln | f'(x) | \rangle_S = \ln 2$. Therefore, in the limit of large $N$, the observable has zero mean, that is, $\langle A^*\rangle \to 2 \ell - \ln 4 = 0$.

\begin{figure}
    \centering
    \includegraphics[width=\linewidth]{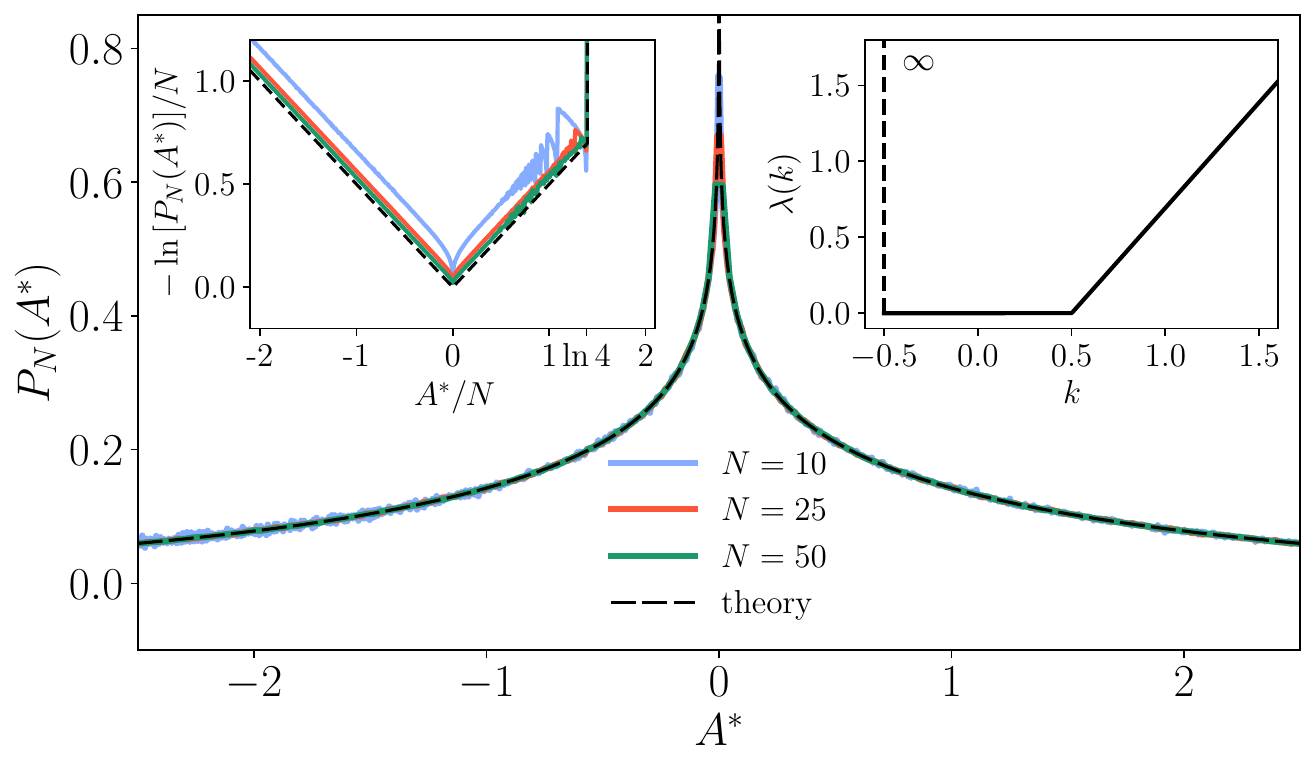}
    \caption{Statistics of the finite‑time Lyapunov exponent (FTLE) $\ell_N$ for the logistic map.  
    In the main panel: PDF of the shifted FTLE $A^*=2\ell_N-N\ln4$ for $N=10$, $25$, and $50$ (colored lines, light to darker).
The dashed black line is the large-$N$ analytic prediction $P_N(A^*) \simeq \tfrac{2}{\pi^{2}} \mathrm{arctanh} \left( e^{-\frac{|A^*|}{2}} \right)$, Eq.~(\ref{LogisticSol}).  
Left inset: convergence of the finite $N$ numerical rate function, $-\ln P_N(A^*)/N$
to the theoretical prediction \eqref{IaSolFTLE}. Note the upper cutoff at $A^*_{\mathrm{max}}=N\ln4$.
Right inset: the SCGF $\lambda(k)$ see Eq.~\eqref{lambdaSolFTLE} (solid black line), exhibiting a non‑analytic point at $k=1/2$ that signals a point of dynamical phase transition. On the other limit, at $k=-1/2$ we have that the SCGF diverges.
 \label{fig:FTLE} }
\end{figure}

For typical fluctuations of $A^*$, or explicitly for $A^* \sim O(1)$, we can use Eq.\,(\ref{eq:PNA}) to find the PDF as
\begin{equation}
\label{LogisticSol}
    P_N(A^*) \simeq \frac{2}{\pi^2} \mathrm{arctanh} \left( e^{-\frac{|A^*|}{2}} \right) \, ,
\end{equation}
which we checked numerically in Fig.~\ref{fig:FTLE}, for moderately large values of $N$.
Eq.~\eqref{LogisticSol} is in perfect agreement with the result that was obtained in \cite{Anteneodo2004} using a rather involved analysis of the large-$N$ behavior of an exact expression for $P_N(A^*)$.
Importantly, we are thus able to explain the properties mentioned above -- namely, the mirror symmetry of the distribution of the FTLE and the anomalous scaling of its variance with $N$ -- as immediate consequences of the fact that $A^*$ is a derived observable. An additional property that follows from this fact, is that (for the logistic map) the FTLE, evaluated on any periodic orbit, exactly equals the Lyapunov exponent [since $A^*=0$ for any periodic orbit, see Eq.~\eqref{eq:periodic} above].

Let us now discuss large deviations of the FTLE in the logistic map.
The SCGF and the rate function that describe the distribution of the FTLE have been known for some time, obtained through a direct calculation of the moment generating function \cite{Grassberger1988}. Through the appropriate shift and rescaling, Eq.~(\ref{eq:scaled-FTLE}), we can immediately deduce those of our observable $A^*$:
\begin{eqnarray}
\label{lambdaSolFTLE}
        \lambda(k) &=&
\begin{cases}
\infty & \text{if } k < -\tfrac{1}{2} \\[1mm]
0 & \text{if } -\tfrac{1}{2} < k < \tfrac{1}{2} \\[1mm]
(2k-1) \ln 2 & \text{if } \tfrac{1}{2} < k
\end{cases} \, ,\\[1mm]
\label{IaSolFTLE}
    I(a) &=&
\begin{cases}
\frac{|a|}{2} & \text{if } a < \ln 4 \\[1mm]
\infty & \text{if } a \geq \ln 4
\end{cases} \, ,
\end{eqnarray}
see Fig.~\ref{fig:FTLE}.
Note that the rate function's behavior around $a=0$ matches smoothly with the $A^* \gg 1$ tail of the formula \eqref{LogisticSol} for the typical-fluctuations regime.
We will now provide the form of the left and right eigenfunctions of the tilted Frobenius-Perron operator for $|k| < 1/2$ and for positive integer values $k=1,2,\dots$. 

For $|k| < 1/2$, the right eigenfunction corresponding to the largest eigenvalue, $\lambda(k)=0$, can be found immediately from Eq.~(\ref{eq:derived-eigenfunctions}):
\begin{equation}
\label{psiSolLogistic1}
    \psi_R(x) \propto x^k (1-x)^{k-1/2} \, .
\end{equation}
It can only be normalized for $k>-1/2$, with the normalization integral $\int_0^1 \psi_R(x) dx$ diverging otherwise. In the range $k<-1/2$, the mean over the IM $\langle e^{kg(x)} \rangle_S$ diverges, which, from the definition of the SCGF in Eq.\,(\ref{eq:SCGF}) indicates that $\lambda(k)$ also diverges. This property is already known in the literature \cite{Grassberger1988}.

For values of $k>1/2$, although the eigenfunction \eqref{psiSolLogistic1} is normalizable, a different eigenvalue $\lambda(k) = (2 k - 1) \ln 2$, becomes the largest. This means that the system undergoes a dynamical phase transition at the critical value $k=1/2$. We have been able to find analytically the eigenfunctions for positive integer $k$, by applying the ansatz
\begin{equation}
    \psi_R(x) \propto \frac{(1-x)^{k/2}}{\sqrt{x}} q_k(x) \, ,
\end{equation}
where $q_k(x)$ is a polynomial of order $k-1$. The coefficients of the polynomial are found by plugging this ansatz into the eigenfunction equation \eqref{eq:tilted-operator-logistical} [with $g(x)$ given by \eqref{gStarDefFTLE}]  and matching the coefficients of the powers of $x$ on both sides of the equation.
The solutions for the first few integer $k$'s are
\begin{subequations}
\begin{align}
    q_1(x) &= 1 \, , \\
    q_2(x) &= 1 - \tfrac{2}{3}x \, , \\
    q_3(x) &= 1 - x + \tfrac{2}{15} x^2 \, . 
\end{align}
\end{subequations}
These solutions are associated with the eigenvalue $\lambda(k) = (2k - 1 )\ln 2$.
For noninteger values $k>1/2$, we were unable to obtain the right eigenfunction analytically, and we leave this as a possible direction for future research.

 The rate function diverges at $a > \ln 4$, which, in particular, indicates that the approximate mirror symmetry of $P_N(A^*)$ breaks down in the large-deviations regime. Let us now explain these two features using the representation of $A^*$ as a derived observable.
The observable can be written exactly as
\begin{equation} 
A^* = \ln x_{N+2} - \ln x_2 \, . \end{equation} 
To obtain large negative values of $A^*$, one must have $x_{N+2} \to 0$ while keeping $x_2$ of order unity. This restriction is easily met, even in the large deviations regime, since the logistic map maps numbers $x \simeq 1$ to $f(x) \simeq 0$.

On the other hand, obtaining large positive values of $A^*$ requires $x_2$ to be very small.
For sufficiently small values $x_2 \ll 4^{-N}$, $x_{N+2}$ can no longer be considered to be
independent of $x_2$, and satisfies $x_{N+2} \sim 4^N x_2$, leading to
\begin{equation} 
A^* \simeq \ln(4^N x_2) - \ln x_2 = N \ln 4 \, . 
\end{equation} 
Therefore, we can have any arbitrarily large negative values of $A^*$, while large positive values are bounded from above by $A^* = N \ln 4$ (this is indeed an upper bound, since $x_{n+1} \le 4 x_n$ for any $n$, and therefore $x_{N+2} \le 4^N x_2$).

\section{Final Remarks \label{sec:final-remarks}}

We have shown that seemingly distinct integrated observables of chaotic maps can share a large deviation rate function. This striking similarity arises when the difference between the two observables considered takes the form of a derived function [i.e., it can be written in the form \eqref{eq:definition-derived-function}]. As a result, the observables differ only by boundary terms that do not scale with the length $N$ of the chaotic sequence. Entire families of observables, obtained by adding a derived function to the original observable function, share the same rate functions.
We demonstrated these properties for different observables in the tent and logistic maps.

Next, we studied integrated observables that are given by the sum of a given derived function $g^*(x)$ applied to the elements of the chaotic sequence. We showed that the distribution of typical fluctuations of such observables becomes independent of $N$ at $N \gg 1$, and approaches a limiting mirror-symmetric form which we are able to calculate, and which is in general not Gaussian.
Large deviations of such observables display trivial behavior if $g^*(x)$ is bounded, but can behave nontrivially if $g^*(x)$ is unbounded.

\new{For a specific piecewise-linear map, modeling random walkers, we found that the position is a derived observable and the motion is localized and not diffusive (i.e., the effective diffusion coefficient vanishes). We also analytically calculated the rate function that describes the dynamical activity, and showed that it also describes the total number of hops in a single direction (right or left). Further, as another important particular example,} we showed that the FTLE in the logistic map is a derived observable (after applying a proper shift and rescaling). This enabled us to reproduce previously known results for this example in a simple way. 
\new{Recent works have uncovered similarities in deterministic many-body chaotic systems, notably in Floquet cellular automata, where the large deviation statistics of several space-time observables follow the same rate functions~\cite{Klobas2024,DeFazio2024,Garrahan_privcom_2025}. This similarity is due to telescoping cancellations, equivalent to the derived-observable mechanism demonstrated here.} { These examples show that derived observables, despite being non-generic, are of importance in a broad class of dynamical systems.}

Our study opens the path to several interesting possible directions for future work. It would be interesting to try to characterize the class of derived functions. In particular, it would be useful to develop a simple method to test whether a given function $g(x)$ is derived, and if so, to find the corresponding function $h(x)$ for which Eq.~\eqref{eq:definition-derived-function} is satisfied. \new{For all the examples in this manuscript, we managed to obtain the $h(x)$ functions by trial and error.}
Moreover, it would be interesting to search for similar phenomena in continuous-time dynamical (chaotic) systems. This could have important implications for the homogenization method, in which fast chaotic dynamics is replaced by white noise \cite{Just2001, Kelly2017, Lucarini2023}. In particular, the diffusion coefficient predicted by the homogenization method is expected to vanish in cases that are analogous to the ``derived observables'' which we studied here.

\section*{Acknowledgments}

NRS acknowledges useful discussions with Noam Neer, Matan Tal and Assaf Katz.
We acknowledge support from the Israel Science Foundation (ISF) through Grant No. 2651/23. NRS acknowledges support from the Golda Meir fellowship.

\bigskip

\appendix

\section{Details of calculations of the SCGF and the rate function \label{appendix:analytical-continuation}}

\subsection{ Quasi-analytic continuation of the power series}

\begin{figure}
    \centering
    \includegraphics[width=0.9\linewidth]{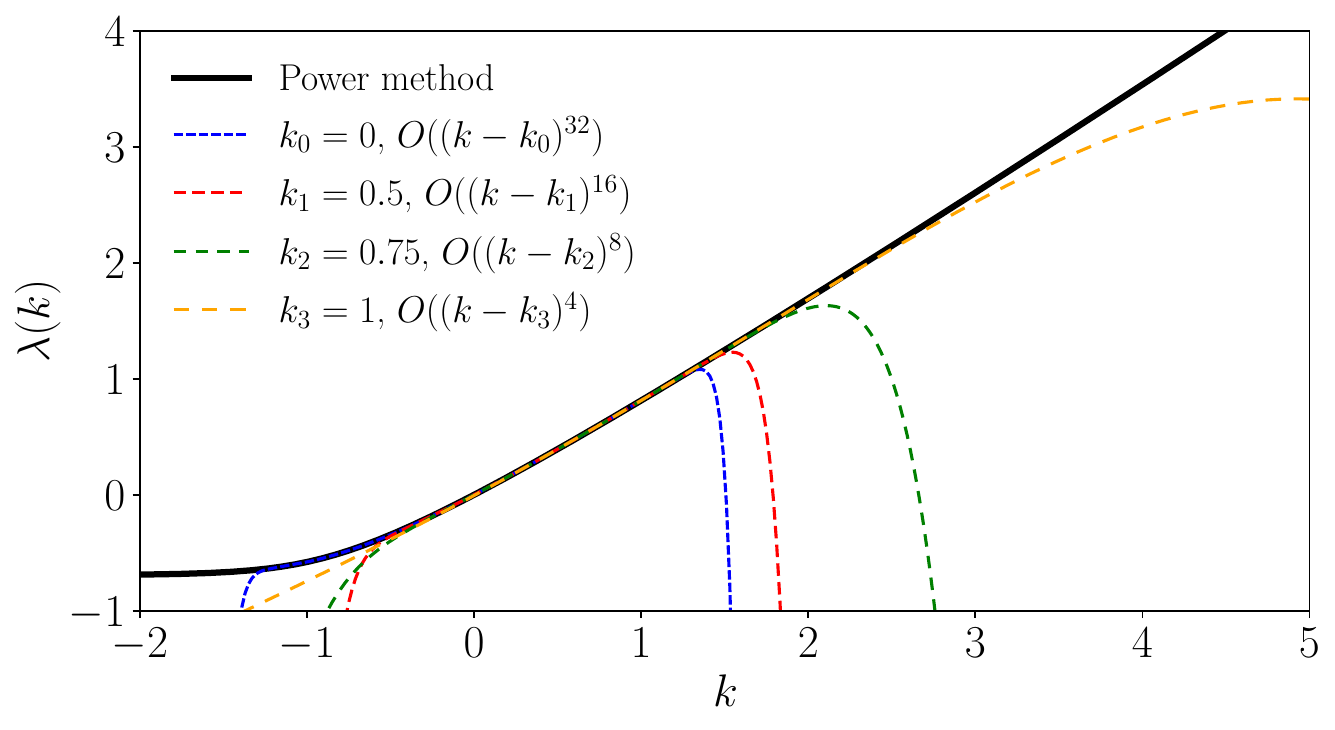}
    \caption{Quasi-analytic continuation of the SCGF $\lambda(k)$ for the tent map. 
The black curve shows the benchmark values of the scaled cumulant generating function $\lambda(k)$ obtained with the power method.
Coloured dashed lines correspond to truncated Taylor series obtained from the recursion, re‑expanded about four successive centers: $k_0=0$ to ${O}((k-k_0)^{32})$, $k_1=0.5$ to ${O}[(k-k_1)^{16}]$, $k_2=0.75$ to ${O}[(k-k_2)^{8}]$, and $k_3=1$ to ${O}[(k-k_3)^{4}]$. 
The excellent overlap between successive local expansions extends the original small‑$k$ solution well beyond its convergence radius $k_c\!\approx\!1.3$, providing an accurate representation of $\lambda(k)$ up to $-1\lesssim k\lesssim 1$ and showing the validity of the quasi-analytical continuation.}
    \label{fig:lambda-numerics}
\end{figure}

The SCGF $\lambda(k)$ can be obtained through a perturbative solution of the tilted eigenvalue problem in the small‑$k$ limit \cite{Smith2022}. This is followed by a quasi-analytic continuation that pushes the result beyond the convergence radius for the original expansion, which we introduce here.  The idea is to write the right eigenfunction as a power series of $k$ as
\begin{equation}
    \psi_R(x) = p_S(x) \left[ 1 + k \, \phi_1(x) + k^2 \phi_2(x) + \cdots  \right] \, ,
\end{equation}
where $\phi_i(x)$ are an ansatz and $p_S(x)$ is the IM. The SCGF is also expanded as
\begin{equation}
    \lambda(k) = k \, \lambda_1 + k^2 \lambda_2 + \cdots \, .
\end{equation}
These two expansions are then plugged into the eigenvalue equation $L_k \psi_R(x) = e^{\lambda(k)} \psi_R(x)$ (Eq.\,(\ref{eq:tilted-operator})).
This procedure yields functional equations for each of the $\phi_i$'s that are simpler and do not involve $k$.
For the tent map $f(x)=1-|1-2x|$ and $g(x)=\frac{3}{2}x$, the ansatz functions $\phi_i(x)$ are polynomials of degree $i$ \cite{Smith2022}, which we write the first three orders as
\begin{subequations}
\begin{align}
    \phi_1(x) &= \tfrac32\,x,\\
    \phi_2(x) &= \tfrac32\,x^{2}-\tfrac34\,x,\\
    \phi_3(x) &= \tfrac98\,x^{3}-\tfrac{21}{16}\,x^{2}+\tfrac38\,x,
\end{align}
\end{subequations}
and
\begin{equation}
    \lambda(k) \;=\;
        \tfrac{3}{4}\,k + \tfrac{3}{32}\,k^{2} - \tfrac{3}{64}\,k^{3} + \mathcal{O}(k^{4}).
    \label{eq:tent_lambda_series}
\end{equation}
This procedure may be carried out explicitly to arbitrary order in $k$ for the tent and logistic maps, with $g(x) \propto x$, see \cite{Smith2022}. Indeed, for the purpose of plotting Fig.~\ref{fig:lambda-numerics}, we carried it out to the 32nd order (with the help of Mathematica \cite{Mathematica}).
However, although $\lambda(k)$ exists (and is finite) for all real $k$, we find empirically that the radius of convergence of the power series \eqref{eq:tent_lambda_series} is finite, and this limits the range of $k$'s for which this series can be used to directly evaluate $\lambda(k)$.

In order to overcome this difficulty, it is desirable to expand $\lambda(k)$ around points other than $k=0$. We developed a method to do so, directly from the series expansion \eqref{eq:tent_lambda_series} around $k=0$, as we now explain.
Let us assume that we have obtained the truncated series centered at $k_0$ of order $M_0$, which we denote $\tilde{\lambda}_{M_0}(k,k_0)$.
We may then perform a Taylor expansion of this truncated series at a different point $k_1$ of order $M_1$, closer to the edge of the convergence radius,
\begin{equation}
    \tilde{\lambda}_{M_1}(k,k_1) = \sum_{m=0}^{M_1}
        \frac{\lambda^{(m)}_{M_0}(k_1,k_0)}{m!}\,(k-k_1)^{m} \, .
    \label{eq:local_taylor}
\end{equation}
If $M_1 = M_0$, then both series are identical, but for $M_1 < M_0$, we can produce a new power series that is valid for values of $k$ beyond he convergence radius of the original series. We can perform the same Taylor series procedure again, now near a point $k_2>k_1$ with $M_2 < M_1$.
 As long as the $M_i$'s are all sufficiently large, this procedure continues the original series \eqref{eq:tent_lambda_series} quite well, and in the limit where $M_i$'s are all infinite, this continuation of the series becomes an analytic continuation.
In other words, the true $\lambda(k)$ can be obtained for any $k$ to arbitrary precision using this method.
 The validity of this procedure is shown in Fig.~\ref{fig:lambda-numerics} for the tent map.

The rate function is obtained as follows. We first apply the Legendre-Fenchel transform \eqref{eq:legendre-fenchel} to the series $\tilde{\lambda}_{M_0}(k,k_0)$. For the tent and logistic maps, this reduces to the Legendre transform \cite{Smith2022}
\begin{equation}
a = \lambda'(k), \quad I(a) = ak-\lambda(k) \, .
\end{equation}
This results in a power series expansion of the rate function $I(a)$ around its minimum $a=a^*$.
However, the radius of convergence of this power series does not cover the entire domain of $I(a)$.
We therefore apply the quasi-analytic continuation method, and this indeed yields $I(a)$ on its entire domain.
This method is one of the two methods that we used to calculate the theoretical curves for $I(a)$ in Fig.~\ref{fig:dynamical_observable_power} of the main text.

\subsection{Power method for calculating $\lambda(k)$ \label{app:power-method}}

Since the analytical approach requires us to calculate the series expansion to very high terms, we also present an alternative direct semi-analytic route to the largest eigenvalue of the tilted operator $L_k$, Eq.\,(\ref{eq:tilted-operator}). We will make use of the standard power method for any value of $k$. Both approaches yield identical results and either one can be used to produce the theoretical lines in Fig.\,\ref{fig:dynamical_observable_power}.

Since $e^{\lambda(k)}$ is the largest eigenvalue of $L_k$, repeated application of $L_k$ on the initial density (the IM in our case) converges exponentially to the right eigenfunction $\psi_R(\bm{x})$, while the growth factor between consecutive iterates converges to $e^{\lambda(k)}$. As mentioned, we use the IM $p_S(x)$ of the map as a starting density; each iteration consists of \begin{equation}
    \phi_{n+1}(x) = L_k \phi_{n}(x) \, ,
\end{equation}
with $\phi_{1}(x) = p_S(x)$. The numerical estimate is found using the expression
\begin{equation}
    e^{\lambda(k)} \approx
    \frac{\int_0^1 \phi_{n+1}(x)\,dx}{\int_0^1 \phi_{n}(x)\,dx} \, ,
    \label{eq:lambda_power}
\end{equation}
for a sufficiently large choice of $n$, with the error estimation being $\mathrm{error}_n = |\frac{\int_0^1 \phi_{n+1}(x)\,dx}{\int_0^1 \phi_{n}(x)\,dx} -  \frac{\int_0^1 \phi_{n}(x)\,dx}{\int_0^1 \phi_{n-1}(x)\,dx}|$. We continue iterating until $\mathrm{error}_n < 10^{-12}$.

Numerically, for the tent-map, the densities $\phi_{n}(x)$ are expressed at positions in a uniform grid
$x_i=i/N_\mathrm{grid}$ with $\Delta x=1/N_\mathrm{grid}$. For the logistic map, the divergence at the edges of $p_S(x)$ is circumvented by working with a new variable $\theta$, defined through $x = 1 - \cos(\theta)$, so that $d\theta = p_S(x) dx$. Therefore, for the logistic map, we use a uniformly distributed grid in $\theta_i = 2 \pi ( i / N_\mathrm{grid})$.
Because the used tent and logistic maps are not invertible, each image point $x_i$ possesses two pre‑images $z_{\pm}=f^{-1}(x_i)$, neither of which generally coincides with a grid node.  We therefore evaluate
\begin{equation}
    L_k\phi(x_i)
    \;=\;
    \frac{1}{2}\,
    e^{k\,g(z_{+})}\,
    \big[\,\phi\big](z_{+}) + \frac{1}{2}\,
    e^{k\,g(z_{-})} \,
    \big[\,\phi\big](z_{-}),
\end{equation}
where $\big[\phi\big](x_i)$ denotes a cubic‐spline interpolation, which reduces the discretisation error to $\mathcal{O}(\Delta x^4)$ of the vector $\{\phi(x_i)\}$. This is called the Ulam method of discretization \cite{Schutte2016}.

\section{Monte Carlo Simulations \label{appendix:MC}}

In this appendix, we describe the Monte Carlo (MC)  simulation approach employed for calculating the probability distribution function (PDF) of an observable $A$, defined by
\begin{equation}
A = \sum_{n=1}^N g(x_n) \, .
\end{equation}
We specifically consider two illustrative chaotic maps: the tent map and the logistic map in Eqs.\,(\ref{eq:def-logistic}) and (\ref{eq:def-tent}). Naively, one obtains the PDF $P_N(A)$ numerically by sampling an initial condition $x_1$ from the IM $p_S(x_1)$ and then evolving it forward to generate a deterministic chaotic sequence up to $x_N$. However, this forward simulation approach is very inefficient for sampling the rare values of $A$.

We use an alternative strategy \cite{Smith2022} by initiating the simulation from the final position $x_N$, also sampled from the IM, and reconstructing the chaotic sequence backward. This backward evolution, unlike the forward process, is not deterministic since the mappings $f(x)$ are not invertible, thus resulting in a Markovian chain. At each backward step, we have two possible pre-images $y_{\pm}$ satisfying $f(y_{\pm}) = x_{n+1}$. Due to the symmetry of $f(x)$, each branch $y_{\pm}$ is selected with probability $1/2$.

Generating the sequence backwards enables us to bias the dynamics towards atypical values of $A$.
For a sequence size $N$, each backward-generated sequence is characterized by $N_1$ choices from the upper branch $y_+$ and $N_0 \equiv N - N_1 - 1$ choices from the lower branch $y_-$. The full probability distribution $P_N(A)$ is thus obtained by conditioning on $N_1$ according to the law of total probability:
\begin{equation}
P_N(A) = \sum_{N_1 = 0}^{N-1} P_N(A|N_1) \mathrm{Prob}(N_1) \, ,
\end{equation}
where $P_N(A|N_1)$ denotes the conditional PDF of $A$ for trajectories with exactly $N_1$ upper-branch selections. The probability $\mathrm{Prob}(N_1)$ is given by a binomial distribution
\begin{equation}
\label{PNAN1}
\mathrm{Prob}(N_1) = \frac{1}{2^{N-1}} \frac{(N-1)!}{N_1!(N - N_1 - 1)!} \, ,
\end{equation}
and we compute the conditional distribution $P_N(A|N_1)$ numerically.

The advantage of this backward MC method lies in its efficient sampling of rare events: extreme values of the observable $A$ correlate strongly with trajectories characterized by extreme values of $N_1$. This allows the method to direct computational resources towards the most relevant regions of the distribution.

\begin{figure}
    \centering
    \includegraphics[width=\linewidth]{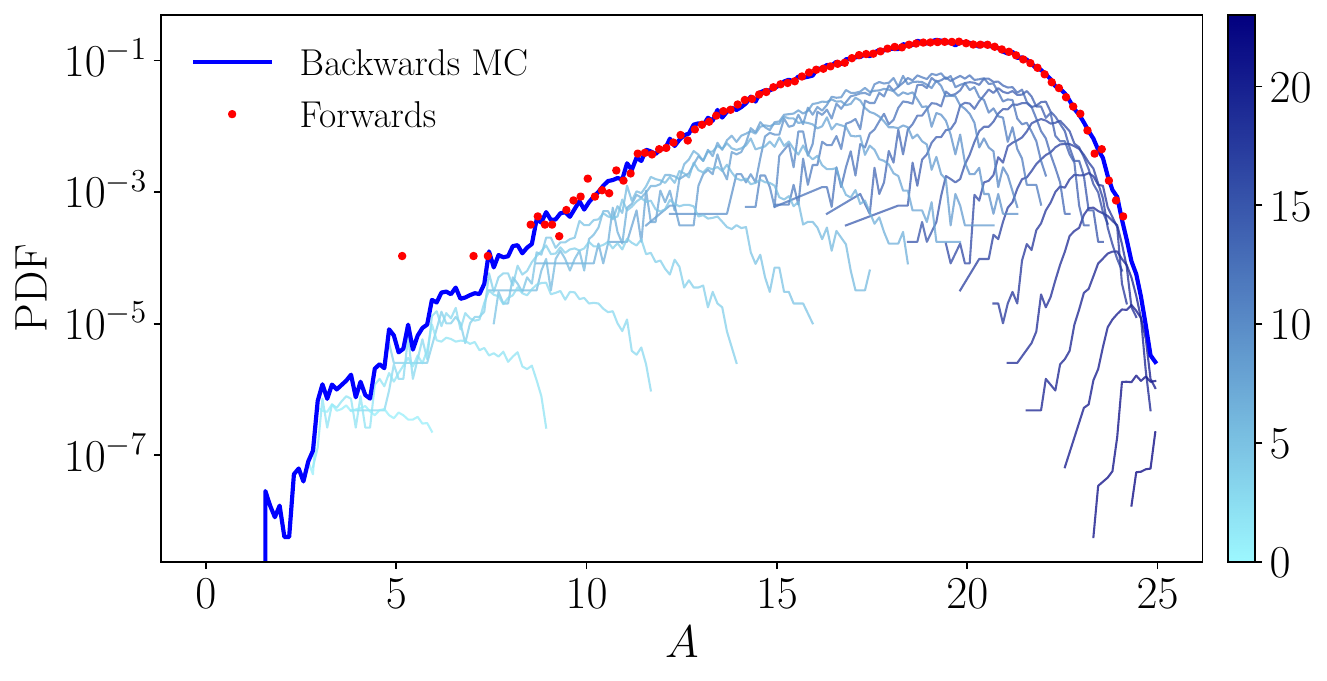}
    \caption{The terms $P_N(A|N_1) \mathrm{Prob}(N_1)$ that appear in the sum \eqref{PNAN1} for the tent map with $N=25$ and observable $g(x)=3x/2$. Thin lines represent different values of $N_1$, the number of upper-branch selections during backward iteration, with colors indicating $N_1$ according to the gradient bar. The thick solid blue line is the sum of these terms obtained from the backward MC simulation method using $2000$ samples per condition. Red points show results from the standard forward simulation using $50000$ samples. These sampling numbers were chosen to be small in order to emphasize the effectiveness of the backward MC simulation. 
    }

    \label{fig:MC-numerics-demonstration}
\end{figure}

Figure \ref{fig:MC-numerics-demonstration} illustrates the conditional PDFs $P_N(A|N_1)$ obtained for the tent map using $N=25$ and $g(x)=3x/2$, based on $2000$ samples per condition. These conditional distributions exhibit minimal overlap and clearly demonstrate convergence to the theoretical large deviation prediction. In comparison, the forward simulation method fails to adequately sample the rare-event tails of the distribution, even when employing a significantly larger number of samples, $50000$, approximately equal to the total number of samples used in the backward-biased approach.

{
\section{Derivation of Eqs.~\eqref{eq:relationRight} and \eqref{eq:relationLeft}}
\label{app:eigenvectorRelation}

As mentioned in the main text, the product of the left and right eigenvectors for $g_1$ and $g_2$ is expected to be the same, i.e.,
\begin{equation}
\label{eigenvectorProduct}
\psi_{R}^{(1)}(\bm{x})\psi_{L}^{(1)}(\bm{x}) = \psi_{R}^{(2)}(\bm{x})\psi_{L}^{(2)}(\bm{x}) \, .
\end{equation}
It is therefore sufficient to derive one of the two equations \eqref{eq:relationRight} and \eqref{eq:relationLeft}, and the other one is expected to follow from the relation \eqref{eigenvectorProduct}. 

Let us now derive Eq.~\eqref{eq:relationLeft}.
Our starting point is the eigenvalue equation \eqref{Lkdagger} for the left eigenvectors for $g_1$ and $g(2)$ (note that we are using the fact that the SCGF $\lambda(k)$ is the same for both observables)
\begin{align}
    e^{kg_{1}(\bm{x})}\psi_{L}^{\left(1\right)}(f(\bm{x}))&=e^{\lambda(k)}\psi_{L}^{\left(1\right)}(\bm{x}) \, ,\\
    e^{kg_{2}(\bm{x})}\psi_{L}^{\left(2\right)}(f(\bm{x}))&=e^{\lambda(k)}\psi_{L}^{\left(2\right)}(\bm{x}) \, .
\end{align}
Dividing the first equation by the second, and multiplying the result by $e^{-kh\left(\bm{x}\right)}$, we obtain
\begin{equation}
\label{psi1divpsi2}
    e^{-kh\left(f\left(\bm{x}\right)\right)}\frac{\psi_{L}^{\left(1\right)}\left(f(\bm{x})\right)}{\psi_{L}^{\left(2\right)}\left(f(\bm{x})\right)}=e^{-kh\left(\bm{x}\right)}\frac{\psi_{L}^{\left(1\right)}\left(\bm{x}\right)}{\psi_{L}^{\left(2\right)}\left(\bm{x}\right)} \, .
\end{equation}
We now notice that in Eq.~\eqref{psi1divpsi2}, both sides of the equation involve the evaluation of the same function at the two points $f(\bm{x})$ and $\bm{x}$ respectively, i.e., if one denotes the right-hand side of \eqref{psi1divpsi2} by
$\xi\left(\bm{x}\right)=e^{-kh(\bm{x})}\frac{\psi_{L}^{\left(1\right)}\left(\bm{x}\right)}{\psi_{L}^{\left(2\right)}\left(\bm{x}\right)}$
then the equation simply becomes
\begin{equation}
\xi\left(f\left(\bm{x}\right)\right)=\xi\left(\bm{x}\right) \, .
\end{equation}
This equation is trivial to solve; its solution is a constant function $\xi\left(\bm{x}\right) = \text{const}$. Arbitrarily choosing this constant to be unity (since the eigenfunctions are anyway only defined up to multiplication by a constant) and recalling the definition of $\xi\left(\bm{x}\right)$, one obtains Eq.~\eqref{eq:relationLeft} as required.
}

{
\section{Calculating the rate function of the activity of the RW via Legendre-Fenchel transform}
\label{app:Legendre}

We begin from the formula \eqref{eq:eigenvalue-diffusion} for the SCGF. Since it is differentiable and convex, its Legendre-Fenchel transform reduces to its Legendre transform \cite{TouchetteLF2014}.
Taking the derivative of \eqref{eq:eigenvalue-diffusion}, we find
\begin{equation}
    a = \frac{1}{2}-\frac{1}{2\sqrt{4e^{k}+1}} \, ,
\end{equation}
which we invert to obtain
\begin{equation}
    k = \ln\left(\frac{1}{4\left(1-2a\right)^{2}}-\frac{1}{4}\right) \, .
\end{equation}
Integrating the latter equation, with the initial condition $I(a_*) = 0$ where $a_* = a|_{k=0} = 1/2 - 1/(2\sqrt{5})$, we obtain the formula for $I(a)$ that is given in Eq.~\eqref{IofaRW} of the main text.
}

\end{document}